\documentstyle[prd,aps,epsfig,preprint,tighten]{revtex}
\begin{document}
\preprint{                                                 BARI-TH/260-97}
\draft
\title{Earth regeneration effect in solar neutrino oscillations:\\
	                  an analytic approach}
\author{          Eligio Lisi and Daniele Montanino}
\address{     Dipartimento di Fisica and Sezione INFN di Bari,\\
                  Via Amendola 173, I-70126 Bari, Italy}
\maketitle
\begin{abstract}
We present a simple and accurate method for computing analytically the 
regeneration probability of solar neutrinos in the Earth. We apply this 
method to the calculation of several solar model independent quantities 
than can be measured by the SuperKamiokande and Sudbury Neutrino Observatory 
experiments.
\end{abstract}

\section{Introduction}

	The Mikheyev-Smirnov-Wolfenstein (MSW) mechanism of neutrino 
oscillations in matter \cite{MSWm} represents a fascinating solution to 
the long-standing solar neutrino problem \cite{NuAs}. The possible 
observation of the $\nu_e$ regeneration effect in the Earth \cite{olds} 
would be a spectacular, solar model independent confirmation of this 
theory (for reviews, see \cite{Ku89,Mi89}).

	The available data from the real-time solar neutrino experiment 
at Kamioka \cite{Hi91,Fu96} are consistent with no Earth regeneration 
effect within the quoted uncertainties. This information can be used to 
exclude a region of the neutrino mass-mixing parameters in fits to the 
solar neutrino data \cite{Ha93,Fi94,Fo96}.

	A larger region of oscillation parameters relevant for the Earth 
effect will be probed by the new generation of solar neutrino experiments
(as shown, e.g., in \cite{Ba94,Kr96}). In particular, the SuperKamiokande 
experiment \cite{To95} (running) and the Sudbury Neutrino Observatory (SNO) 
experiment \cite{Do96} (in construction) are expected to probe possible 
day-night modulations of the solar neutrino flux with unprecedented 
statistics and accuracy. A correct interpretation of the forthcoming 
high-quality data will demand precision calculations of the Earth-related 
observables.

	The calculation of the Sun-Earth $\nu_e$ survival probability 
$P_{SE}(\nu_e)$ is based on the relation (Mikheyev and Smirnov, in 
\cite{olds})
\begin{equation}
P_{SE}(\nu_e)=P_S(\nu_e)+
\frac
                         {[2P_S(\nu_e)-1]\,[\sin^2\theta-P_E(\nu_2\to\nu_e)]}
                                           {\cos2\theta}
\ ,
\label{eq:P_SE}
\end{equation}
where $P_S(\nu_e)$ is the $\nu_e$ survival probability at the Earth surface
(or daytime probability), and $P_E(\nu_2\to\nu_e)$ is the probability of 
the transition from the mass state $\nu_2$ to $\nu_e$ along the neutrino 
path in the Earth.%
\footnote{The derivation of Eq.~(\protect\ref{eq:P_SE}) is reported
in Appendix~A for completeness.}

	The calculation of $P_E$ is notoriously difficult. Since the electron 
density in the Earth is not a  simple function of the radius, the MSW 
equations have  to be  integrated numerically, unless step-wise approximations 
are adopted at the price of lower precision. Moreover, $P_E$ must 
be averaged over given intervals of time,
\begin{equation}
\langle P_E\rangle=
\frac
         	{\displaystyle\int^{\tau_{d_2}}_{\tau_{d_1}}d\tau_d
          	\int^{\tau_{h_2}(\tau_d)}_{\tau_{h_1}(\tau_d)}d\tau_h\,
         	P_E(\eta(\tau_d,\,\tau_h))}
         	{\displaystyle\int^{\tau_{d_2}}_{\tau_{d_1}}d\tau_d
		\int^{\tau_{h_2}(\tau_d)}_{\tau_{h_1}(\tau_d)}d\tau_h}
\ ,
\label{eq:PDH}
\end{equation}
where $\tau_d$ and $\tau_h$ are the daily and hourly times, respectively,
and $\eta$ is the nadir angle of the sun at the detector site. In typical
applications, the interval $[\tau_{d_1},\,\tau_{d_2}]$ covers one year and
the intervals $[\tau_{h_1}(\tau_d),\,\tau_{h_2}(\tau_d)]$ 
cover the nights, but other choices are possible.

	The integration in Eq.~(\ref{eq:PDH}) is time-consuming. For 
instance, the authors of Refs.~\cite{Ha93} and \cite{Fi94} quote a grid of 
about $30\times 30$ integration points in the ${\rm (year)}\times{\rm (night)}$ 
domain, which requires massive calculations for spanning the relevant region of
neutrino mass and mixing parameters with acceptable precision. The issue
of numerical accuracy and stability is not secondary, since coarser 
integrations may generate fuzzy and misleading results (see, e.g., Fig.~5 of 
\cite{Ga95}).

	A  faster and more elegant method for averaging $P_E$ consists in 
transforming the double integral of Eq.~(\ref{eq:PDH}) into a single integral 
of the form
\begin{equation}
\langle P_E\rangle = 
		\displaystyle\int^{\eta_2}_{\eta_1}d\eta\,W(\eta)\,P_E(\eta)
\ ,
\label{eq:PETA}
\end{equation}
where the weight function $W(\eta)$ represents the ``solar exposure'' of
the trajectory corresponding to the nadir angle $\eta$. This method was used
by Cherry and Lande \cite{olds} for calculating the day-night asymmetry 
at the Homestake site. We have used this approach in our previous
works \cite{Fo96} by computing numerically the Jacobian $d\tau/d\eta$
required to transform Eq.~(\ref{eq:PDH}) into Eq.~(\ref{eq:PETA}).

	In this paper we show that, actually, the weight function $W(\eta)$
can be calculated analytically in several cases of practical interest.
Moreover, we show that $P_E$ can  also be calculated analytically through 
a simple approximation which is more accurate than is required 
by the present (imperfect) knowledge of the Earth's interior.
Within this approach, we work out the calculation of several solar
model independent observables for the SuperKamiokande and SNO experiments
in a two-family oscillation scenario.

	Our work is organized as follows. In Sec.~II we present the 
parametrization of the electron density. In Secs.~III and IV we discuss the 
analytic calculation of $P_E$ and $W$, respectively. In Sec.~V we apply 
these calculations to the SuperKamiokande and SNO experiments. We draw our 
conclusions in Sec.~VI. In order to make this work as self-contained and 
useful as possible, we organize in Appendixes~A--C the relevant mathematical 
proofs. The reader interested mainly in the final results for  
SuperKamiokande and SNO may skip Secs.~II--IV and the appendixes, 
and read only Sec.~V.

\section{Parametrizing the Earth electron density}

	In solar neutrino physics, the ``standard electroweak model'' of 
particle physics and the ``standard solar model'' of astrophysics must be 
supplemented by a ``standard Earth model'' of geophysics, such as the 
Preliminary Earth Reference Model (PREM) of Anderson and Dziewonsky 
\cite{PREM}. This seemingly ``preliminary''  model elaborated in 1981
still represents the standard framework for the interpretation of 
seismological data,%
\footnote{For a review of recent progresses in the study of the Earth interior,
see also \cite{Re95}.}
as far as possible shell asphericities are
neglected \cite{ASPH}.

	Eight shells are identified in the PREM model, but for any 
practical purpose related to solar neutrinos the four outer shells can be 
grouped into a single one (the ``upper mantle''). The Earth matter 
density profile $\rho(r)$ is given in detail in Table~I of \cite{PREM}.

	We have derived the electron density profile $N(r)$ from $\rho(r)$
by assuming the following chemical compositions (in weight): (1) Mantle,
SiO$_2$ (45.0\%), Al$_2$O$_3$ (3.2\%), FeO (15.7\%), MgO (32.7\%), and
CaO (3.4\%) \cite{MANT}; (2) Core, Fe (96\%) and Ni (4\%) \cite{CORE}. 
It follows that
\begin{equation}
			N/\rho =  
\left\{\begin{array}{cl}
				0.494 &,\;{\rm(mantle)}\ ,\\
				0.466 &,\;{\rm(core)}\ .	
\end{array}\right.
\end{equation}

\begin{figure}
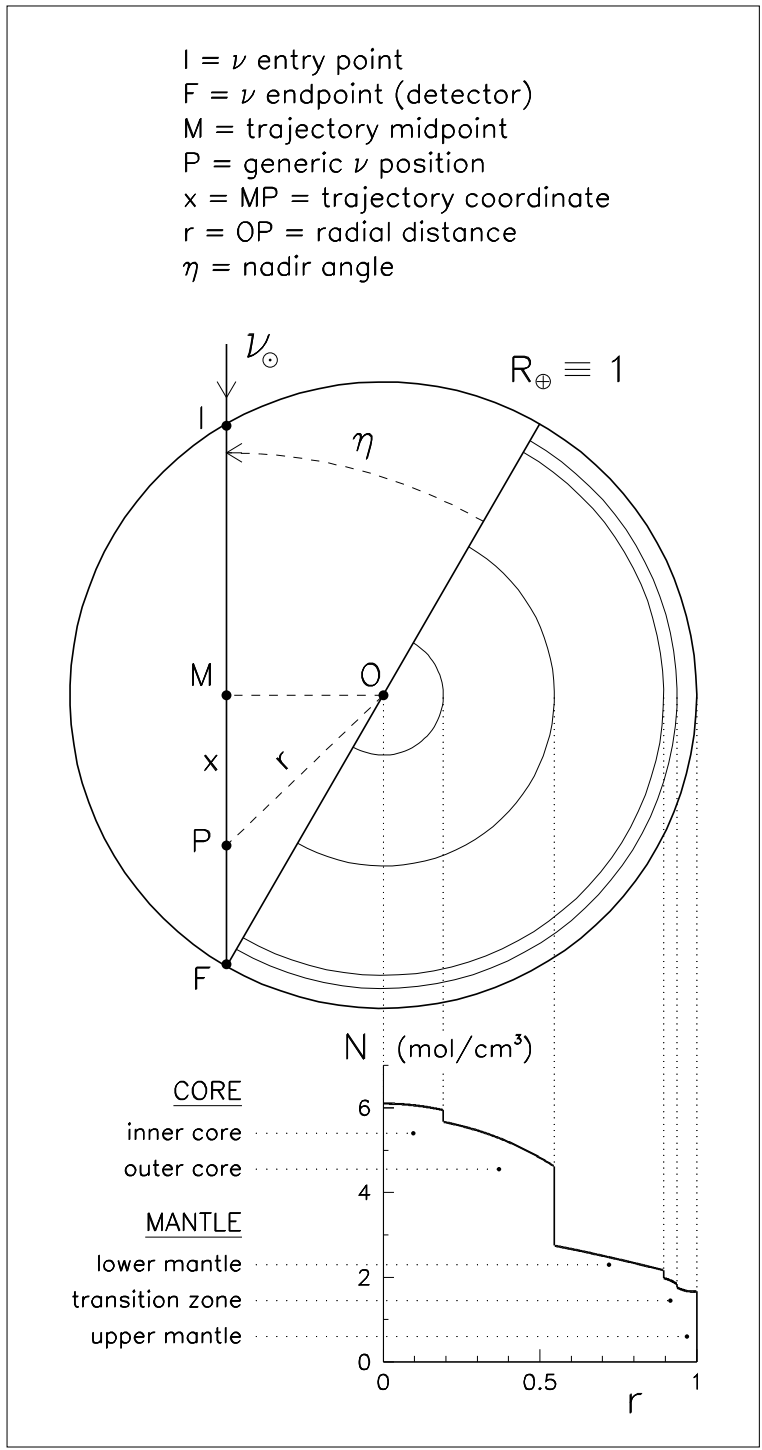

\caption{	Section of the Earth showing the relevant
		shells (in scale) and  the electron density profile 
		$N(r)$. The geometric definitions used in the text 
		are also displayed.}
\label{F:1}
\end{figure}

	Figure~\ref{F:1} shows the five relevant Earth shells (in scale)
and the electron density $N(r)$, together with the basic geometry that will 
be used in the following sections. For each shell $j$, we  use a polynomial 
fit that approximates accurately the true radial density,
\begin{equation}
	N_j(r) = \alpha_j + \beta_j r^2 + \gamma_j r^4
\ ,
\label{eq:Nj(r)}
\end{equation}
where the coefficients $\alpha_j$, $\beta_j$, and $\gamma_j$ 
are given in Table~\ref{tb:N}.

	The functional form in Eq.~(\ref{eq:Nj(r)}) is invariant for 
nonradial $(\eta\neq 0)$ neutrino trajectories:
\begin{equation}
	N_j(x) = \alpha'_j + \beta'_j x^2 + \gamma'_j x^4
\ ,
\label{eq:Nj(x)}
\end{equation}
where
\begin{mathletters}
\begin{eqnarray}
   \alpha'_j &=& \alpha_j +    \beta_j \sin^2\eta + \gamma_j \sin^4\eta\ ,\\
   \beta'_j  &=& \beta_j  + 2 \gamma_j \sin^2\eta                      \ ,\\
   \gamma'_j &=& \gamma_j                                            
\ ,
\label{eq:abc}
\end{eqnarray}
\end{mathletters}
with the trajectory coordinate $x$ and the nadir angle $\eta$ defined
as in Fig.~\ref{F:1}.

	For later purposes it is useful to split the density (in each
shell and for each trajectory) as
\begin{equation}
	N_j(x)={\overline N}_j+\delta N_j(x)
\ ,
\label{eq:Nsplit}
\end{equation}
where $\overline N$ is the ($\eta$-dependent)
average density along the shell chord,
\begin{equation}
	{\overline N}_j=\int_{x_{j-1}}^{x_j}dx\,N_j(x)\bigg/(x_j-x_{j-1})
\ ,
\label{eq:Nbar}
\end{equation}
and  $\delta N_j(x)$ is the residual density variation. It will be seen 
that the above parametrization of $N(x)$ plays a basic role in the 
analytic calculation of the neutrino  probability $P_E$.

	We end this section with an estimate of the likely uncertainties 
affecting $N(x)$. The core, which is usually assumed to be iron-dominated, 
could contain  a large fraction of lighter elements without necessarily 
conflicting with the  seismological data. An example is given by a model of 
core made of Fe (55\%) and FeO (45\%)  \cite{Re95}, which would increase $N$ 
by 0.65\%. Concerning the mantle, alternative chemical compositions (see 
Table~4 in \cite{MANT}) typically reduce $N$ by 1--2~\%. We will evaluate the 
effect of representative density uncertainties by varying $N$ by $+1$\% in the 
core and by $-1.5$\% in the mantle. However, these error estimates might 
be optimistic, according to Birch's old admonition \cite{Bi52}.

\section{Calculating $P_E$ with elementary functions}

In this section we show that the probability $P_E(\nu_2\to\nu_e)$ 
can be accurately approximated by elementary analytic expressions.
We start by observing that $P_E$ can be expressed as:
\begin{equation}
	P_E = |{\cal U}_{ee}\sin\theta+{\cal U}_{e\mu}\cos\theta|^2 
\ ,
\label{eq:PSEU}
\end{equation}
where ${\cal U}$ is the neutrino evolution operator in the
$(\nu_e,\,\nu_\mu)$ flavor basis.

	In the same flavor basis, the MSW Hamiltonian ${\cal H}_j(x)$ along 
the $j$th shell chord traversed by the neutrino is given by
\begin{equation}
{\cal H}_j(x) = \frac{1}{2}
\left(\begin{array}{cc}
	  \sqrt{2} G_F N_j(x) - k\cos2\theta & 	               k\sin2\theta      
\\
          k\sin2\theta &                  k\cos2\theta - \sqrt{2} G_F N_j(x) 
\end{array}\right)
\ ,
\label{eq:H(x)}
\end{equation}
where $k=\delta m^2/2E_\nu$ is the vacuum oscillation wave number,  $N_j(x)$ 
is the  electron density as in Eq.~(\ref{eq:Nj(x)}), and  
$\delta m^2$, $\theta$,  and $E_\nu$ are the neutrino mass square difference, 
mixing angle, and energy, respectively.

	Following Eq.~(\ref{eq:Nsplit}), we split the Hamiltonian into a 
constant matrix plus a perturbation,
\begin{equation}
	{\cal H}_j(x)={\overline{\cal H}}_j+\delta{\cal H}_j(x)
\ ,
\label{eq:Hsplit}
\end{equation}
where ${\overline{\cal H}}_j={\cal H}_j\big|_{N\to\overline N}$, and
$\delta{\cal H}_j(x)={G_F}/\sqrt{2}\,{\rm diag}
[\delta N_j(x),\,-\delta N_j(x)]$. Notice that the unperturbed Hamiltonian
${\overline{\cal H}}_j$ depends on the nadir angle $\eta$ through 
${\overline{N}}_j$.

	We have worked out explicitly, at the first perturbative order,
the evolution operator ${\cal U}_j$ for the $j$th shell chord in the 
flavor basis. The result is:
\begin{eqnarray}
     	{\cal U}_j(x_j,\,x_{j-1}) 
&=&    	e^{-i{\overline{\cal H}}_j(x_j-x_{j-1})}
   	-i\int_{x_{j-1}}^{x_j}dx\,
   	e^{-i{\overline{\cal H}}_j(x_j-x)}
   	\delta{\cal H}_j(x)
   	e^{-i{\overline{\cal H}}_j(x-x_{j-1})} 
   	+ {\cal O}(\delta{\cal H}_j^2)\\
&=& 	\left(\begin{array}{cc}
	c_j+is_j\cos2\bar\theta_m &                -is_j\sin2\bar\theta_m \\
	-is_j\sin2\bar\theta_m &                   c_j-is_j\cos2\bar\theta_m
	\end{array}\right)-\frac{i}{2}\sin 2\bar\theta_m\nonumber\\
&\times&\left(\begin{array}{cc}
	C_j \sin2\bar\theta_m &                C_j\cos2\bar\theta_m - i S_j\\
	C_j\cos2\bar\theta_m + iS_j &                  -C_j \sin2\bar\theta_m
	\end{array}\right) + {\cal O}(\delta{\cal H}_j^2)
\ ,
\label{eq:U}
\end{eqnarray}
where $\bar\theta_m$ is the average mixing angle in matter,
\begin{equation}
	\sin2\bar\theta_m/\sin2\theta = 
	\left[(\cos2\theta-\sqrt{2}G_F{\overline N}_j/k)^2+\sin^22\theta
	\right]^{-\frac{1}{2}}
\ , 
\label{eq:sinm}
\end{equation}
and
\begin{mathletters}
\begin{eqnarray}
	c_j 	&=& 	\cos[\bar k_m(x_j-x_{j-1})/2]\ , \\
	s_j 	&=& 	\sin[\bar k_m(x_j-x_{j-1})/2]\ , \\
	C_j 	&=& 	\sqrt{2}G_F\int^{x_j}_{x_{j-1}}dx\,\delta N_j(x)
        		\cos \bar k_m(x-\bar x)\ ,\\
	S_j 	&=& 	\sqrt{2}G_F\int^{x_j}_{x_{j-1}}dx\,\delta N_j(x)
        		\sin \bar k_m(x-\bar x)
\ ,
\end{eqnarray}
\label{eq:cjsj}
\end{mathletters}
with $\bar k_m=k\sin2\theta/\sin2\bar\theta_m$ (average matter oscillation
wave number) and $\bar x=(x_j+x_{j-1})/2$ (shell chord midpoint).

	The integrals in  Eqs.~(\ref{eq:cjsj}c) and (\ref{eq:cjsj}d)
are elementary, $\delta N_j$ being a (biquadratic) polynomial in $x$  
(see Sec.~II). The property $\int_{x_{j-1}}^{x_j}dx\,\delta N_j(x)=0$,
which follows from Eqs.~(\ref{eq:Nsplit}) and (\ref{eq:Nbar}), is crucial
for obtaining the compact expression of ${\cal U}_j$ in Eq.~(\ref{eq:U}).

	The evolution operator along the total neutrino path $\overline{IF}$
(see Fig.~\ref{F:1}) is simply given by the ordered product of the partial 
evolution operators along the shell chords, 
${\cal U}(x_F,\,x_I)=\prod_j {\cal U}(x_{j},\,x_{j-1})$.
Actually, due the symmetry of the electron density with respect to 
the trajectory midpoint $M$, one needs only to calculate the evolution
operator from $x_M(=0)$ to $x_F(=-x_I)$,
\begin{eqnarray}
	{\cal U}(x_F,\,x_I) &\equiv & {\cal U}(x_F,\,0)\cdot {\cal U}(0,-x_F)
\nonumber\\
	&=& {\cal U}(x_F,\,0) \cdot {\cal U}^T(x_F,\,0) 
\ .
\label{eq:Ufact}
\end{eqnarray}
The proof of the above property is given in Appendix~B.

	So far we have solved analytically the MSW equations in the Earth at
first order in perturbation theory, by expressing the total evolution operator
$\cal U$ in the flavor basis as a product of matrices (one for each shell 
traversed in a semitrajectory) involving only elementary functions. The 
desired probabilities $P_E$ and $P_{SE}$ are then given by Eqs.~(\ref{eq:PSEU}) and 
(\ref{eq:P_SE}), respectively. Now we discuss the accuracy 
of such first-order approximation.

\begin{figure}
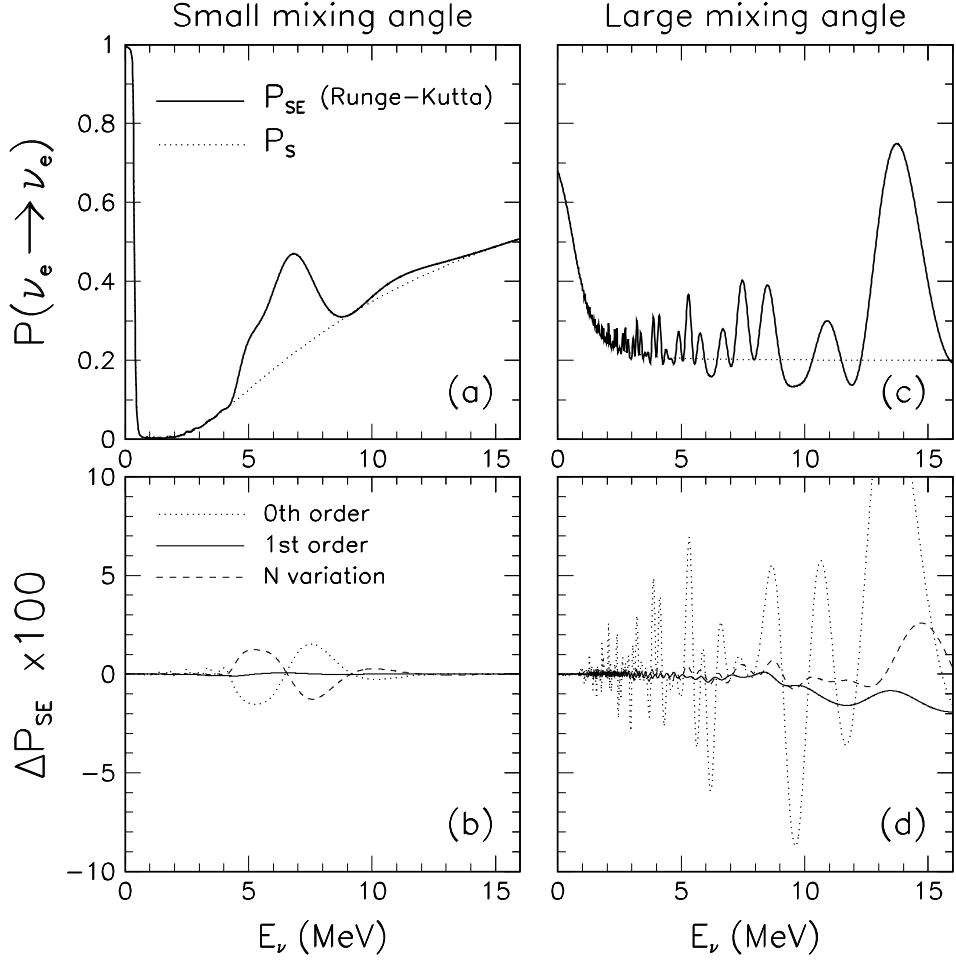

\caption{	Comparison of different calculations of $P_{SE}(E_\nu)$ for
		$^8$B neutrinos crossing the Earth diameter.
		(a) Calculation of $P_{SE}$ with Runge-Kutta integration for
		the small mixing angle case (solid line). 
		Also shown is the function $P_S$ (dotted line).
		(b) Variations of $P_{SE}$ induced by representative
		density shifts (dashed line), and by the first-order
		and zeroth order approximations discussed in the text
		(solid and dotted line, respectively). Panels (c) and (d)
		are analogous to (a) and (b), but refer to the large mixing
		angle case.}
\label{F:2}
\end{figure}

	Figure~\ref{F:2} shows, for two representative mass-mixing scenarios 
and for diametral crossing, the results of various approximations
of $P_{SE}$ as a function of the neutrino energy. 
Figures~\ref{F:2}(a) and \ref{F:2}(b) refer to the small mixing angle
solution to the solar neutrino problem, corresponding to 
$(\delta m^2/{\rm eV}^2,\,\sin^2 2\theta)\simeq(5.2\times 10^{-6},\,8.1\times
10^{-3})$ \cite{Fo96}. In Fig.~\ref{F:2}(a), the probability $P_S$ (dotted 
line)  is calculated semianalytically \cite{Fo96} and averaged over the 
$^8$B production region in the Sun \cite{Ba95} (as required for applications
to SuperKamiokande and SNO). The probability $P_{SE}$ in Fig.~\ref{F:2}(a)
(thick, solid line) has been obtained by integrating  the MSW equations in 
the Earth with the highest possible accuracy,  i.e., with a Runge--Kutta 
method and with the true (PREM) electron density. In Fig.~\ref{F:2}(b) we 
show the residuals $\Delta P_{SE}$ of different calculations with respect to 
the  ``Runge--Kutta''  $P_{SE}$. The solid curve in  Fig.~\ref{F:2}(b)
refers to the first-order perturbative approach discussed in this section. 
The dotted curve is obtained by using the simple zeroth-order approximation 
(i.e., average density shells). The dashed curve shows  the variations of 
$P_{SE}$ induced by plausible uncertainties in the electron density $N$ 
(as discussed at the end of Sec.~II). It can be seen that the effect of the 
latter uncertainties is comparable to the errors associated to the
zeroth-order approximation, and is much larger than the errors of the 
first-order approximation.

	Figures \ref{F:2}(c) and \ref{F:2}(d) are the analogous of 
Figs.~\ref{F:2}(a) and \ref{F:2}(b) for the  large mixing angle solution to 
the solar neutrino problem, corresponding to 
$(\delta m^2/{\rm eV}^2,\,\sin^2 2\theta)\simeq(1.5\times 10^{-5},\,0.64)$ 
\cite{Fo96}. It can be seen that the errors associated to the first-order 
approximation are generally smaller than the effect of the $N$ uncertainties, 
which are in turn much smaller than the errors of the zeroth-order 
approximation.

	The results of Fig.~\ref{F:2} and of many other checks that we have 
performed for different values of $(\delta m^2,\,\sin^2 2\theta)$ and $\eta$
show that the analytic (first-order) solution discussed in this section 
represents a very good approximation to the true electron survival probability
in the Earth, with an accuracy better (often much better) than is
required by the likely  uncertainties affecting the Earth electron density.

	Finally, we point out that our analytic approximation for the neutrino
evolution operator in the Earth matter can be applied also in the analysis of 
atmospheric neutrinos, and that its computer evaluation is much faster 
(about two orders of magnitude) than typical Runge--Kutta numerical 
integrations.

\section{Weighting neutrino trajectories}

	As anticipated in the Introduction, the time average of the neutrino
regeneration probability in the Earth [Eq.~(\ref{eq:PDH})] can be transformed 
into a (more manageable) weighted average over the  trajectory nadir angle 
$\eta$ [Eq.~(\ref{eq:PETA})], with a weight function $W(\eta)$ having a 
compact, analytic form in several cases of practical interest.
In this section we describe the results for the important case of
annual averages during (a fraction of) night. We refer the reader
to Appendix C for mathematical proofs and for a discussion of other cases.

	The weight function $W(\eta)$ for annual averages is presented in 
Table~\ref{tb:W}. In different ranges of the detector latitude $\lambda$ 
and of the nadir angle $\eta$, $W(\eta)$ takes different functional forms, 
involving the calculation of a complete elliptic integral of the first kind
\cite{Gr94,Ab72} (which is coded in many computer libraries; see, e.g.,
\cite{CERN}).

\begin{figure}
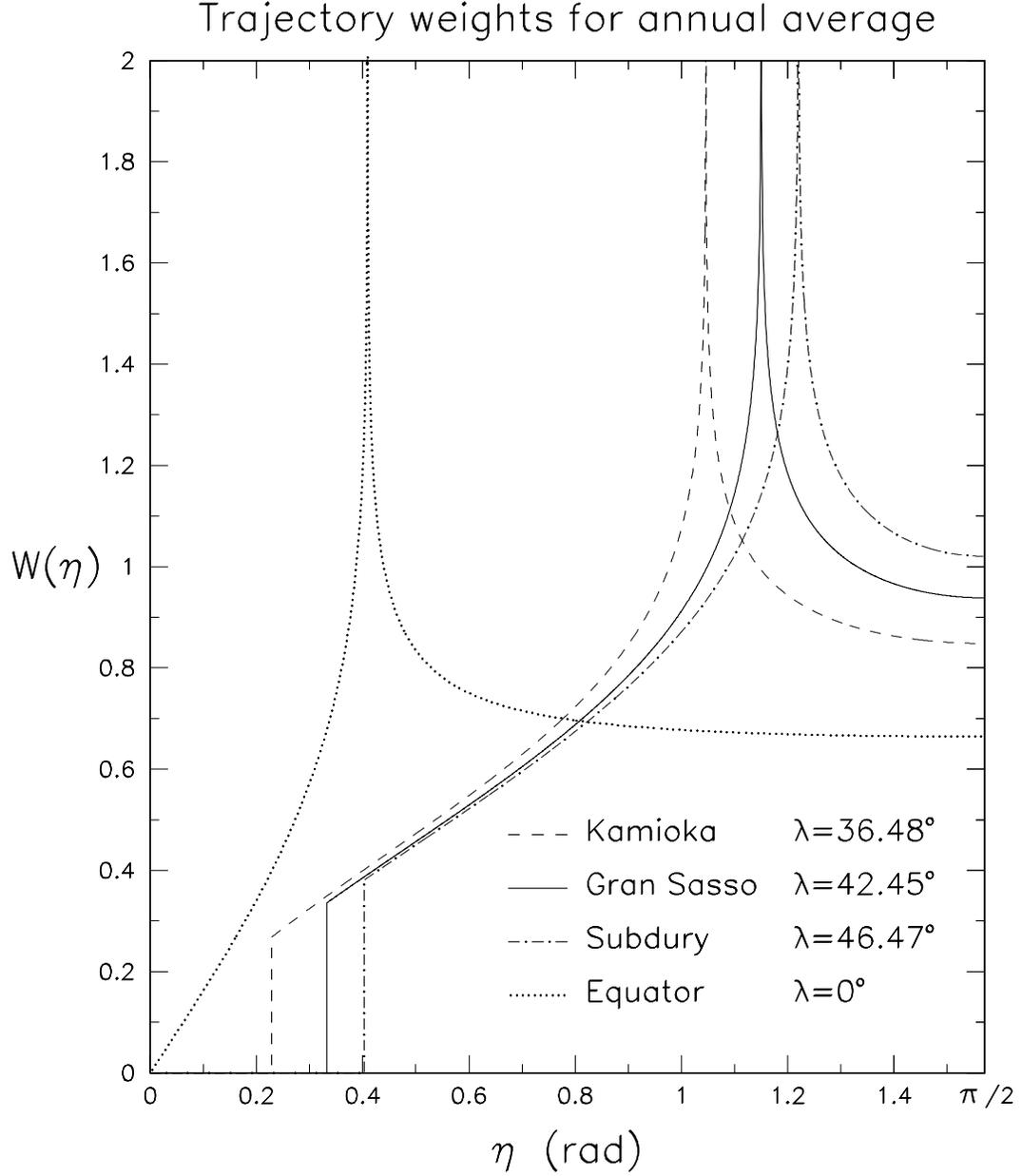

\caption{	Annual solar exposure (weight) of the trajectory
		at nadir angle $\eta$ for representative values
		of the  latitude
		$\lambda$. See the text for details.}
\label{F:3}
\end{figure}

	In Fig.~\ref{F:3} the function $W(\eta)$ is plotted for the
SuperKamiokande and SNO latitudes. The area under each curve is equal to 1. 
We show $W(\eta)$ also for the Gran Sasso site, relevant for several 
proposed solar neutrino projects  such as the Borexino experiment 
\cite{BORE}, the Imaging of Cosmic And Rare  Underground Signals (ICARUS) 
experiment \cite{ICAR},  the permanent Gallium Neutrino Observatory (GNO) 
\cite{PGNO}, and  the Helium at Liquid Azote temperature (HELLAZ) detector
\cite{HELZ}. Finally, the dotted line in Fig.~\ref{F:3} represents
the weight function for a hypothetical detector located at the equator,
where the Earth regeneration effect would be more sizeable \cite{EQUA}.
The divergence of $W(\eta)$ is logarithmic and thus it is integrated
out by binning in $\eta$. Several methods exist for dealing
with the numerical quadrature of divergent integrands \cite{QUAD}.

	By using $W(\eta)$ as given in Table~\ref{tb:W} (or in 
Fig.~\ref{F:3}), the average probability during night simply reads
\begin{equation}
	\langle P_E \rangle_{\rm night} =
	\int^{\pi/2}_0 d\eta\,W(\eta)\,P_E(\eta)
\ .
\label{eq:Pnight}
\end{equation}

	The annual average during the fractions of night in which the Earth 
core is crossed has also a particular relevance as emphasized, e.g., in
\cite{Ba94,EQUA}. With the weight method, it can be easily calculated as
\begin{equation}
	\langle P_E \rangle_{\rm core} =
\frac{\displaystyle
			\int^{\eta_{\rm core}}_0 d\eta\,W(\eta)\,P_E(\eta)}
{\displaystyle 
 			  \int^{\eta_{\rm core}}_0 d\eta\,W(\eta)}
\ ,
\label{eq:Pcore}
\end{equation}
where $\eta_{\rm core}(=0.577$~rad) is the nadir angle subtending the 
Earth (inner and outer) core.

	A final remark is in order. In the expression for the time average 
[Eq.~(\ref{eq:PDH})] we have not included the geometric factor $L^{-2}$ 
accounting for the neutrino flux variations with the Earth-Sun  distance $L$. 
We have implicitly assumed that the data from  the real-time SuperKamiokande 
and SNO experiments will be corrected for this factor in any period of data 
taking. The effect of dropping this assumption is examined in Appendix~D.
We anticipate that, for annual averages, the effect is less than 1\%.

\section{Calculating solar model independent observables}

	The SuperKamiokande and SNO experiments are sensitive only to the 
high-energy part of the solar neutrino spectrum, namely, to $^8$B neutrinos.
Since the estimated uncertainty of the theoretical $^8$B neutrino 
flux $\Phi_B$ is relatively large ($\sim 16\%$ at $1\sigma$ \cite{Ba95}), 
it is important to focus on  observables that do not depend 
on the absolute value of $\Phi_B$, but are 
sensitive only to the {\em shape\/} of the $^8$B energy spectrum (which is 
rather well known \cite{Al96}). Important examples of these quantities are 
the night-day rate asymmetry, the shape distortions of the angular spectrum, 
and the shape distortions of the recoil electron energy spectrum. The SNO
experiment can measure, in addition, the charged-to-neutral current
event ratio, which is also solar model independent. In this Section we 
calculate the annual averages of several such observables, by including the 
Earth effect with the method described in the previous sections.

	We take from \cite{Ba96,TABL} the neutrino interaction cross sections 
for SuperKamiokande. These cross sections already include the effect of 
the detector energy resolution and threshold (see Table~I of \cite{Ba96} for 
the detector technical specifications). Concerning the distortions of the 
electron energy spectra due to neutrino oscillations, we adopt, as in 
\cite{Ba96}, the approach in terms of the first two moments of the electron 
energy distribution, namely, the average  electron kinetic energy 
$\langle T\rangle$ and the variance $\sigma^2$ of the  energy spectrum. 
The reader is referred to \cite{Ba96} for an extensive discussion of the 
spectral moments  and for an estimate of their likely uncertainties.

\subsection{SuperKamiokande}

\begin{figure}
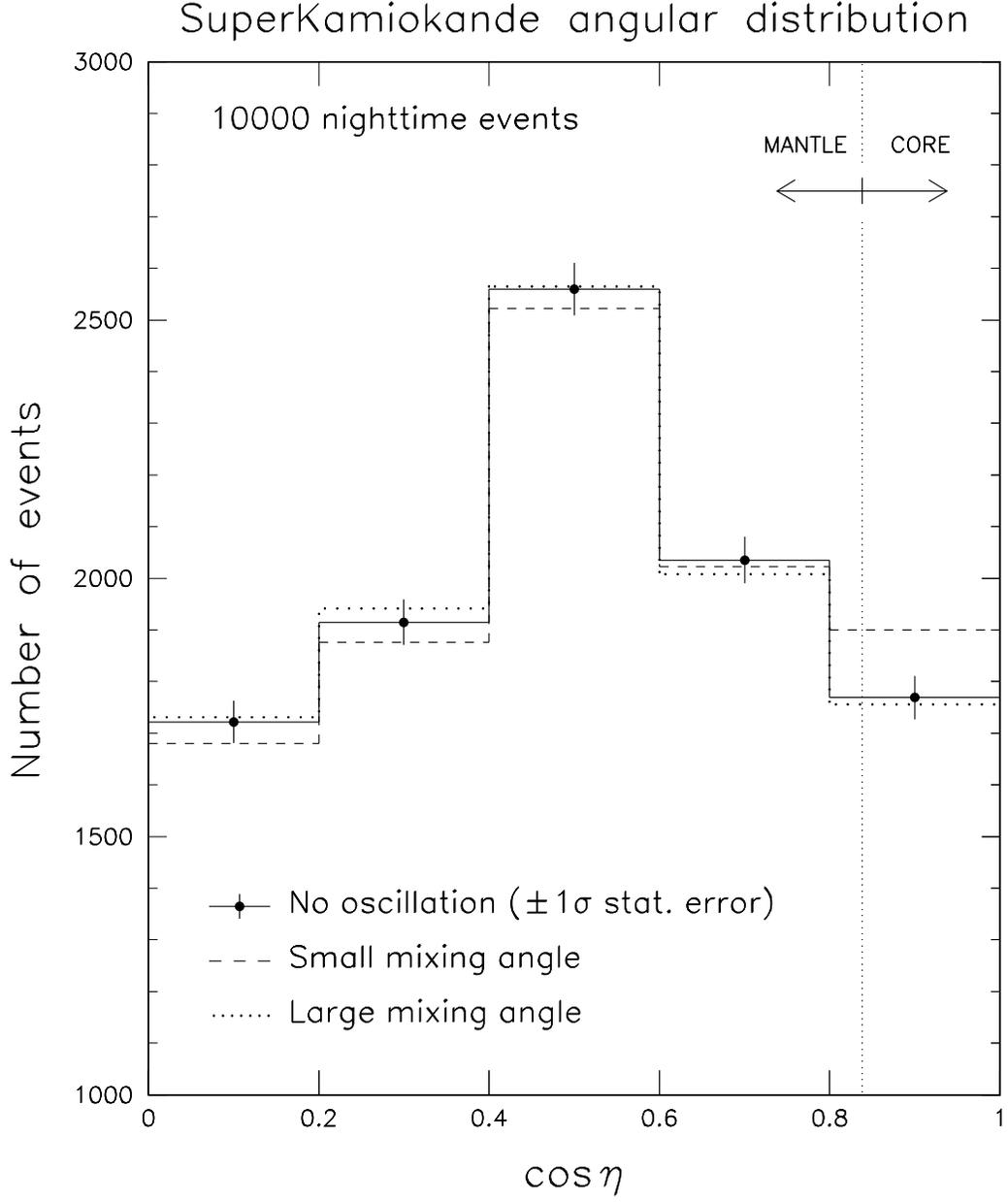

\caption{	Nadir angle distribution of nighttime events at
		SuperKamiokande. Error bars are statistical
		only.}
\label{F:4}
\end{figure}

	Figure~\ref{F:4} shows the nadir angle ($\eta$) distribution of 
events expected at SuperKamiokande. We use the same format (five bins in 
$\cos\eta$) as the Kamiokande experiment \cite{Hi91}. The solid line is 
the distribution expected in the absence of oscillations, which is simply 
obtained by integrating the weight function of Fig.~\ref{F:3} in each bin 
of $\cos\eta$. The dashed and dotted histograms refer to the (best-fit) 
small-mixing and large-mixing solutions, respectively. All histograms are 
normalized to the same number of events in order to make the relative 
deviations independent of the absolute neutrino flux. Assuming a statistics
of 10000 nighttime events, the small mixing angle case appears to be 
separated by $\sim 3\sigma$ (statistical errors only) from the no 
oscillation case in the last bin, which collects neutrinos crossing the 
Earth core. In fact, in the small mixing angle case there is a strong 
regeneration effect in the core (see, e.g., \cite{Ba94}). In the 
(best-fit) large mixing angle case, instead, the sudden variations of 
$P_{SE}$ with $\eta$ \cite{Ba94} happen to be smeared by binning and the 
net deviations are smaller (the effect, however, is very sensitive to 
the specific mass-mixing parameters chosen).

\begin{figure}
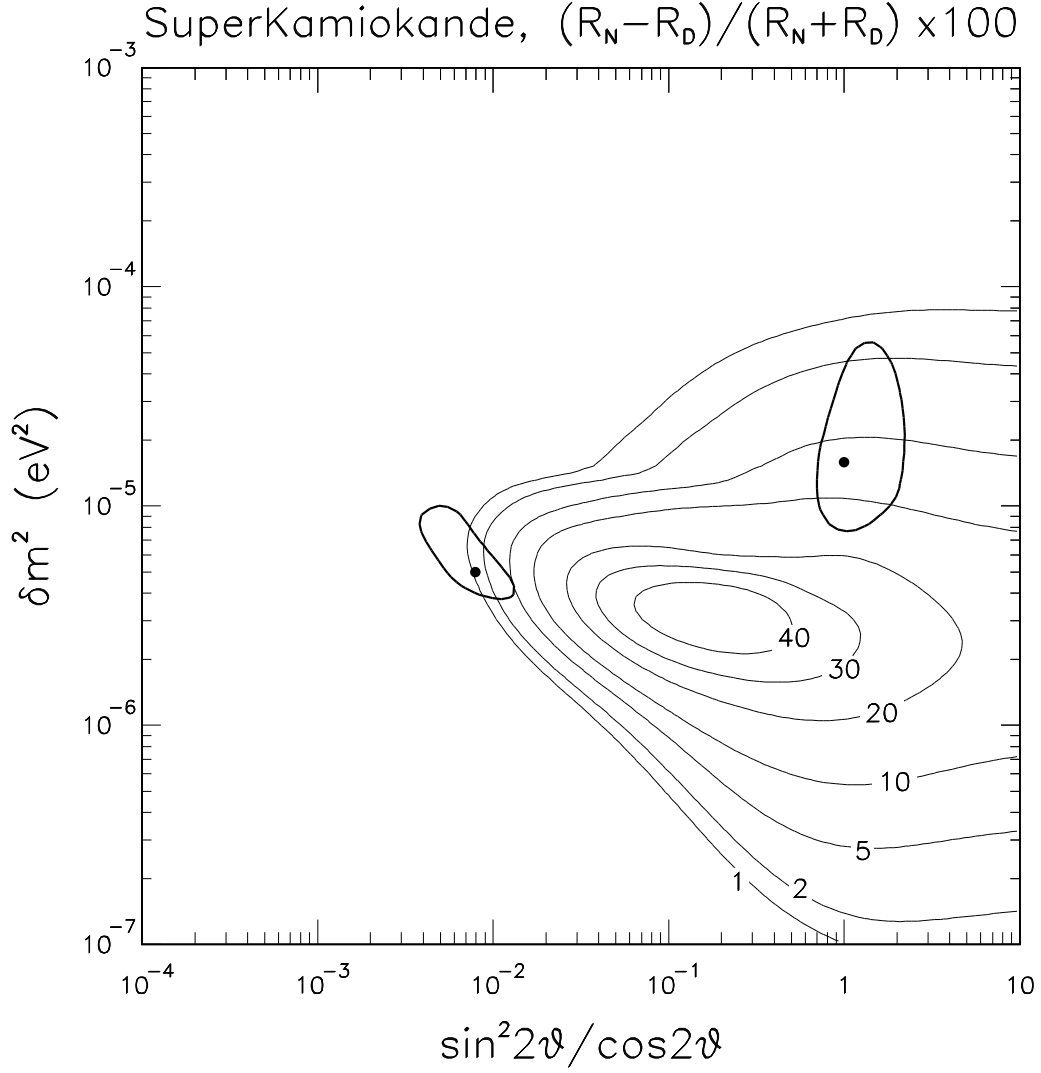

\caption{	Night-day asymmetry of neutrino rates at SuperKamiokande.
		The best-fit points and the 90\% C.L.\ regions
		for the small and large mixing angle solutions are
		superposed.}
\label{F:5}
\end{figure}

	Figure~\ref{F:5} shows the night-day asymmetry of neutrino rates,
which is perhaps the most popular characterization of the Earth effect. 
The 90\% C.L. regions corresponding to the small and large mixing angle 
solution \cite{Fo96} are superposed to curves of equal values of the 
asymmetry. Similar results have been obtained by Krastev in~\cite{Kr96}.
Notice that asymmetry measurements at the percent level would allow
a complete (partial) exploration of the large (small) mixing angle solution.

\begin{figure}
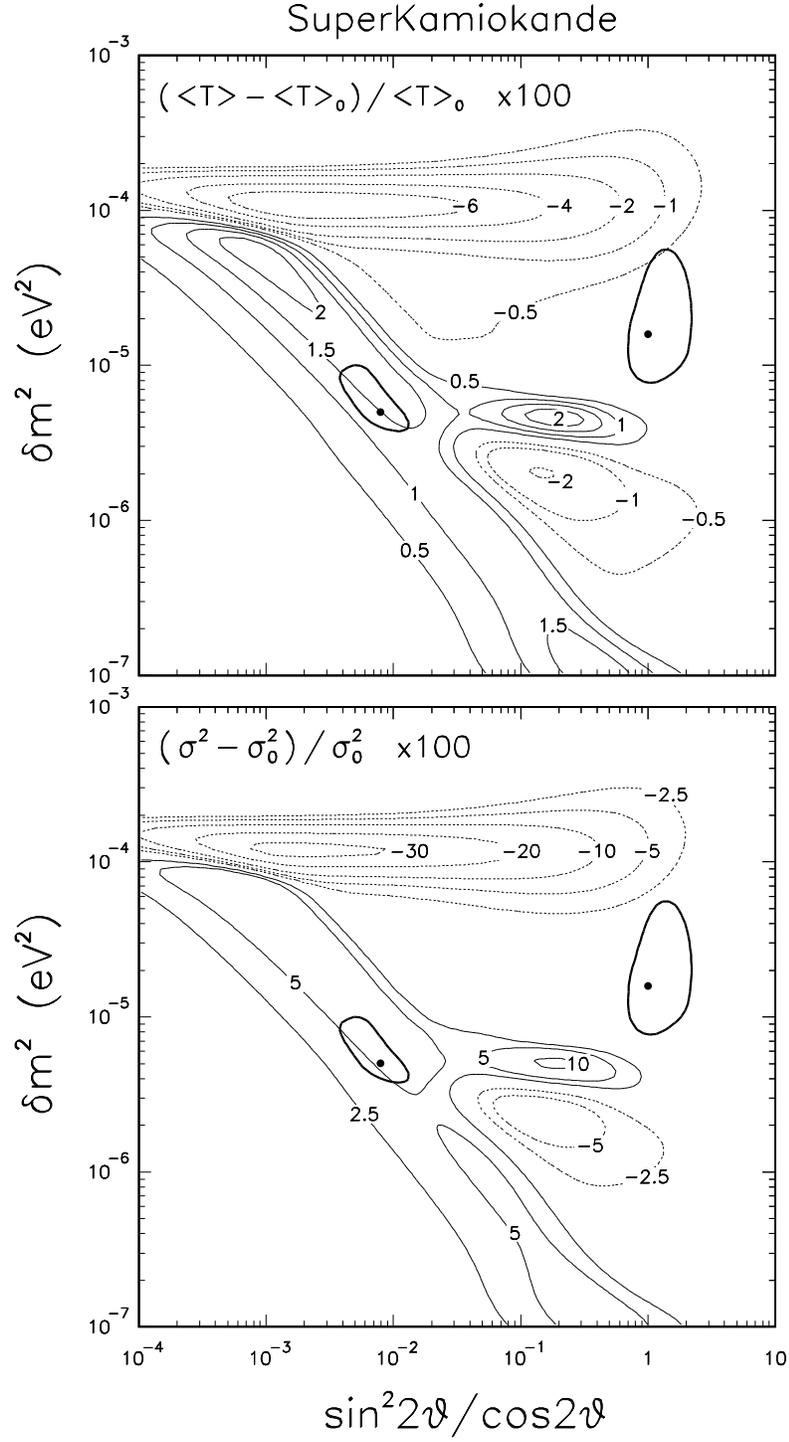

\caption{	Fractional deviations of the first two moments of
		the SuperKamiokande
		electron energy distribution ($\langle T\rangle$
		and $\sigma^2$) from their no-oscillation values
		($\langle T\rangle_0$
		and $\sigma^2_0$).}
\label{F:6}
\end{figure}

	Figure~\ref{F:6} shows the fractional deviations in the first
two moments of the electron energy distribution ($\langle T\rangle$
and $\sigma^2$) with respect to their no-oscillation values 
($\langle T\rangle_0$ and $\sigma^2_0$). These deviations represent a 
useful characterization of the spectral distortions \cite{Ba96}. 
The deviations expected for the small mixing angle solution \cite{Ba96}, 
although significant,  are only slightly affected by the Earth effect. 
The deviations for the large angle solution are very small. In the region 
of ``intermediate'' mixing there could be strong, Earth-related
deformations of the spectrum. Calculations of $\langle T\rangle$ including 
the Earth effect were first presented in \cite{Fi94}. A comparison of their
Fig.~5 \cite{Fi94} with our Fig.~\ref{F:6} shows once again that the accuracy
and stability of numerical calculations of the Earth effect are important
issues. We obtain results very similar to those in Ref.~\cite{Ba96} 
when the Earth effect is switched off.%
\footnote{It is worth mentioning that the computer codes used in this 
work are independent from those used in \protect\cite{Ba96}).}

\begin{figure}
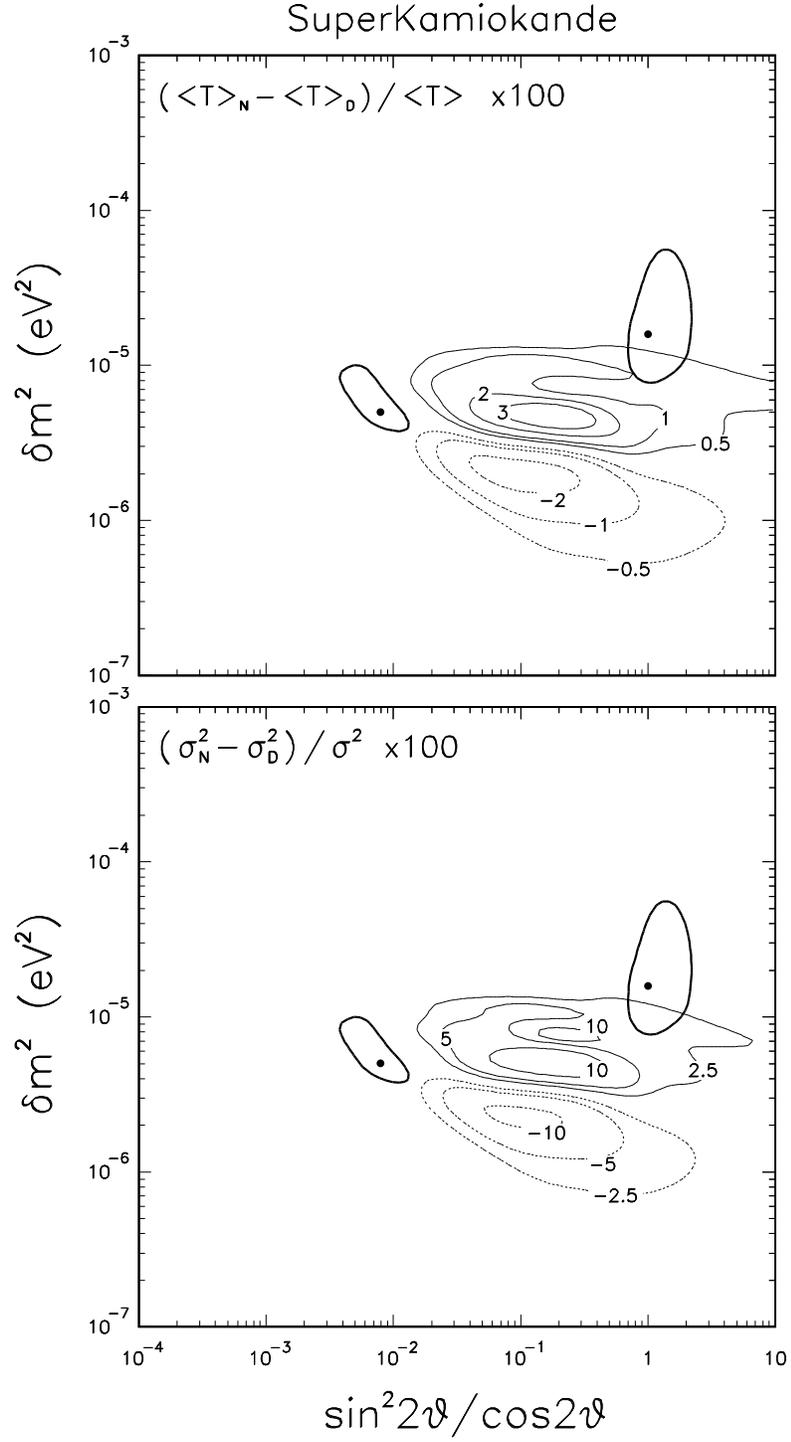

\caption{	Night-day fractional
		variations of the spectral moments at SuperKamiokande.}
\label{F:7}
\end{figure}

	Figure~\ref{F:7} shows the night-day variation of the spectral 
moments at SuperKamiokande. The relative deviations of the nighttime $(N)$ and 
daytime $(D)$ values of $\langle T\rangle$ and $\sigma^2$  characterize the 
daily deformations of the electron spectrum  due to neutrino oscillations in 
the Earth matter (averaged over the year). For $\delta m^2\gtrsim 3\times 10^{-6}$
eV$^2$ ($\delta m^2\lesssim 3\times 10^{-6}$ eV$^2$) the Earth effect tend to
increase the rate in the high-energy (low-energy) part of the electron
spectrum. This explains the sign of the night-day spectral deviations
in Fig.~\ref{F:7}. Therefore, if a significant Earth effect were observed,
the sign of these deviations could provide an additional handle
for discriminating the value of $\delta m^2$.

\subsection{SNO}

\begin{figure}
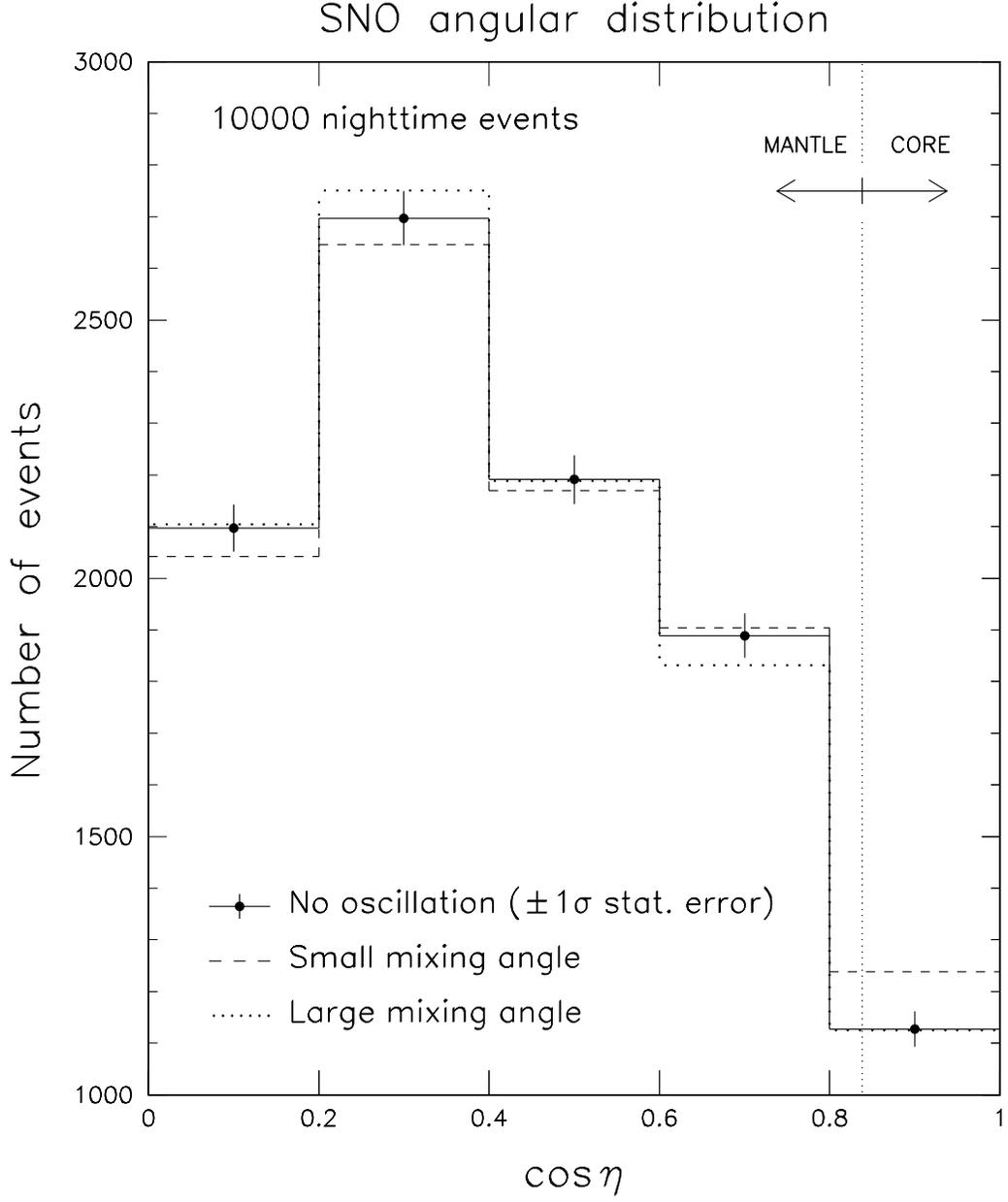

\caption{	Nadir angle distribution of nighttime events at
		SNO. Error bars are statistical
		only.}
\label{F:8}
\end{figure}

\begin{figure}
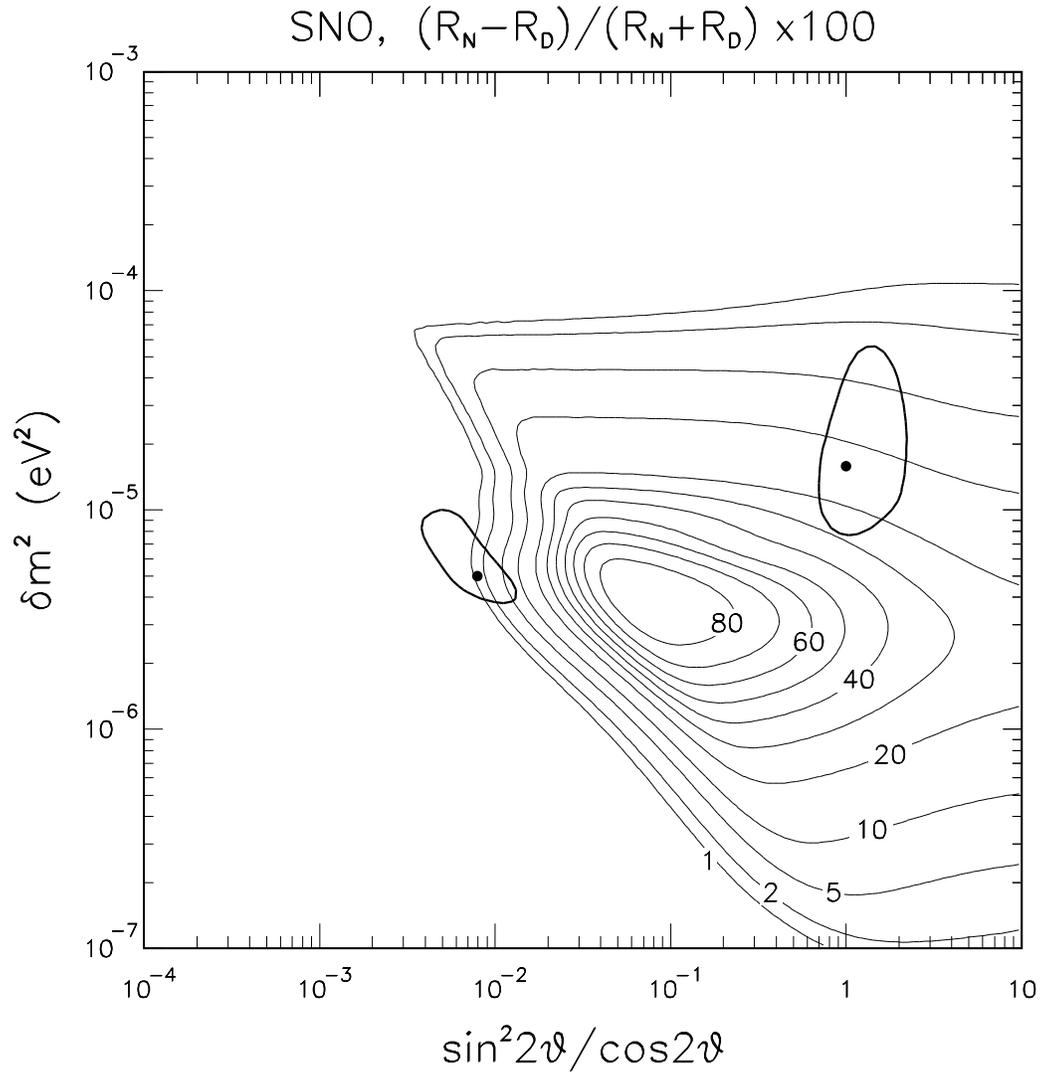

\caption{	Night-day asymmetry of neutrino rates at SNO.}
\label{F:9}
\end{figure}

\begin{figure}
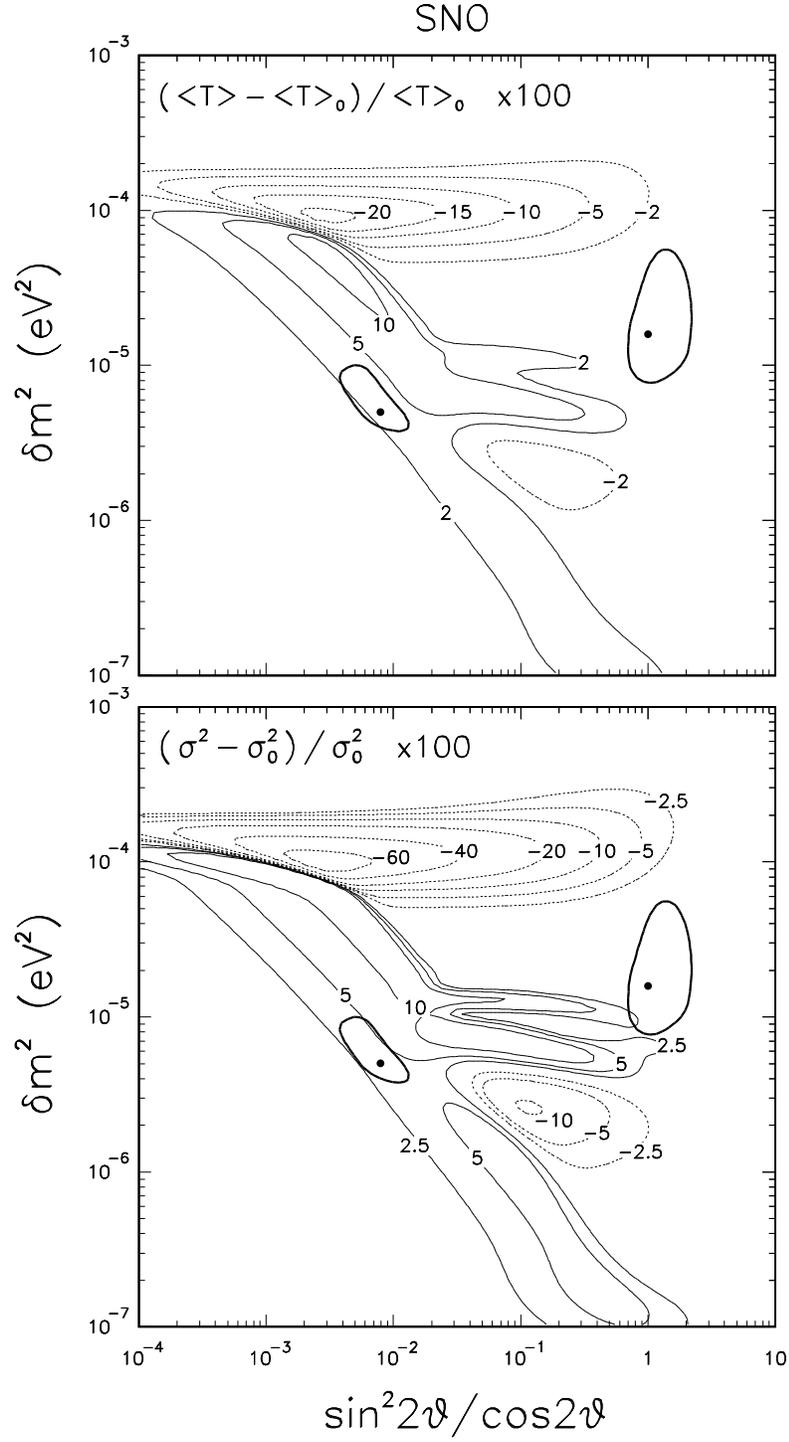

\caption{	Fractional deviations of the first two moments of
		the SNO electron energy distribution ($\langle T\rangle$
		and $\sigma^2$) from their no-oscillation values.}
\label{F:10}
\end{figure}

\begin{figure}
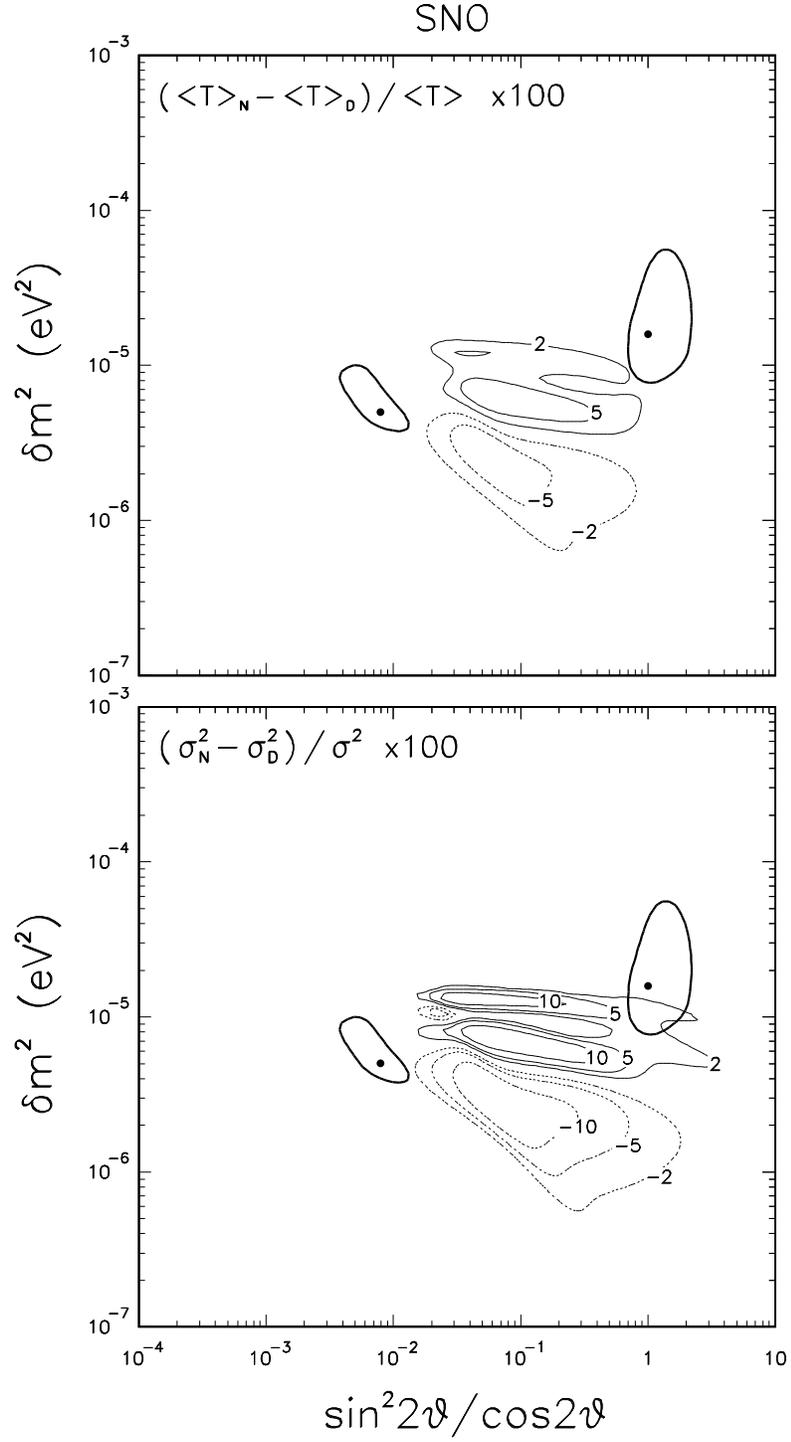

\caption{	Night-day fractional 
		variations of the spectral moments at SNO.}
\label{F:11}
\end{figure}

	The results of our calculation of the angular distribution,
day-night asymmetry, and spectral deviations for the SNO experiments
are presented in Figs.~\ref{F:8}--\ref{F:11}. These figures are analogous 
to Figs.~\ref{F:4}--\ref{F:7} for SuperKamiokande, and similar comments 
apply. We just add that, in general, the various Earth-related effects 
appear to be more significant in SNO than in SuperKamiokande, as a result
of the intrinsically higher correlation between the (observed) 
electron energy and the (unknown) neutrino energy.

\begin{figure}
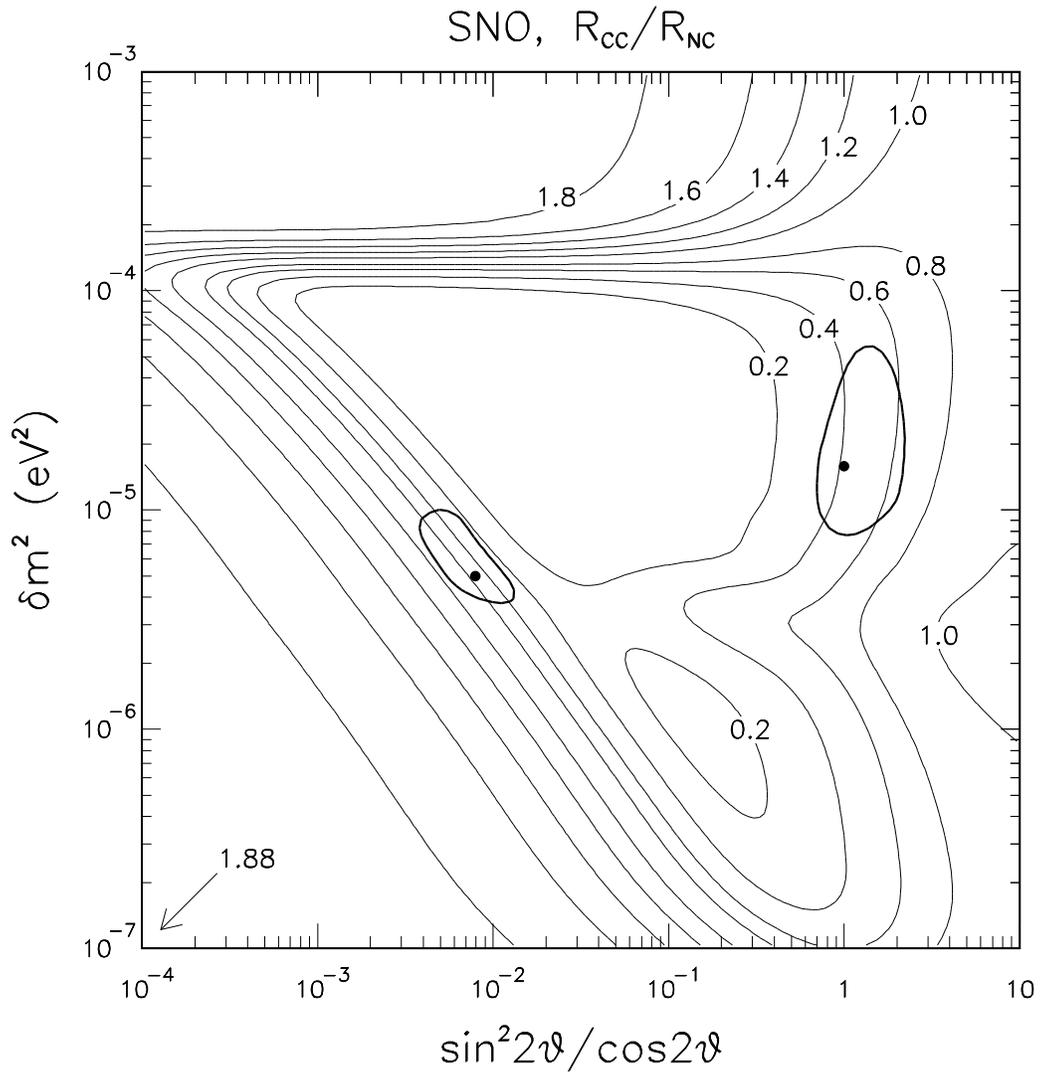

\caption{	Ratio of charged current to neutral current neutrino
		interactions at SNO.}
\label{F:12}
\end{figure}

	In addition, the SNO experiment will separate events produced
in charged current (CC) interactions of $\nu_e$'s from events produced
in neutral current (NC) interactions of neutrinos of all flavors.
The ratio of the CC and NC rates is perhaps the most crucial,
solar model independent observable that will be measured in the
next few years. Curves of the CC/NC ratio, including the Earth
effect,  are shown  in Fig.~\ref{F:12}. The value expected for no
oscillation (indicated in the left, lower corner) agrees with the 
value given in \cite{Ba96}. Notice that we have taken 
the efficiencies for detecting CC and NC events ($\varepsilon_{\rm CC}$
and $\varepsilon_{\rm NC}$, respectively) equal to 100\%. When the true 
experimental efficiencies will be known, the values in Fig.~\ref{F:12} 
should be multiplied by $\varepsilon_{\rm CC}/\varepsilon_{\rm NC}$.

\section{Conclusions}

	The observation of solar neutrino oscillations enhanced by the 
Earth matter would be a spectacular confirmation of the MSW theory. The new 
generation of solar neutrino experiments can probe this possibility with 
unprecedented accuracy. In particular, the interpretation of the forthcoming,
high-quality data from the SuperKamiokande and SNO experiments demands 
precision calculations of the Earth effect in solar neutrino oscillations.

	We have presented an analytic method for approximating the 
$\nu_e$ regeneration probability in the Earth, based on a first-order 
perturbative expansion of the MSW Hamiltonian and on a convenient 
parametrization of the Earth electron density. We have also shown how time 
averages of the $\nu_e$ survival probability can be transformed into 
weighted averages over the nadir angle, with weights that can be 
calculated analytically in several relevant cases. Mathematical proofs 
and final results are described in detail, especially for the case of annual 
averages.

	We have then calculated accurately the following solar model 
independent observables for the  SuperKamiokande and SNO experiments: 
(1) the angular distribution of events; 
(2) the night-day asymmetry of the neutrino rates; 
(3) the fractional deviations of the  first two spectral moments of the 
     electron energy distribution; 
(4) the night-day fractional variations of such moments; and 
(5) the charged-to-neutral current event ratio for SNO.

	The approach to the Earth effect presented in this paper
allows simpler, faster, and  more versatile calculations than brute-force 
integration methods.  We hope that these advantages may lead more people 
to try a do-it-yourself analysis of the Earth regeneration effect
in solar neutrino oscillations.

\acknowledgments

We thank Professor G.\ L.\ Fogli for useful suggestions and for careful
reading of the manuscript. We thank P.\ I.\ Krastev for fruitful discussions.
The work of D.M.\ was supported in part by Ministero dell'Universit\`a
e della Ricerca Scientifica and in part by INFN.

\appendix
\section{The Sun-Earth survival probability $P_{SE}$}

In this appendix we report, for the sake of completeness, the derivation
of Eq.~(\ref{eq:P_SE}) (Mikheyev and Smirnov, in \cite{olds}).
A solar neutrino arriving at Earth in the flavor state $\nu_\alpha$ is an 
incoherent mixture of vacuum mass states $\nu_i$. The corresponding 
probabilities $P_S(\nu_\alpha)$ and $P_S(\nu_i)$ are then given by
\begin{equation}
\left[\begin{array}{c}
	P_{S}(\nu_e)\\ 		
	P_{S}(\nu_\mu)
\end{array}\right]     =
\left[\begin{array}{cc}
				\cos^2\theta&             \sin^2\theta \\
                                \sin^2\theta &               \cos^2\theta 
\end{array}\right]
\left[\begin{array}{c}
	P_{S}(\nu_1)\\ 
	P_{S}(\nu_2)
\end{array}\right]
\ . 
\label{eq:PS}
\end{equation}

	The probability $P_{SE}(\nu_\alpha)$ that a solar neutrino has 
flavor $\alpha$ after traversing the Earth can be expressed as
\begin{equation}
\left[\begin{array}{c}
	P_{SE}(\nu_e)\\ 
	P_{SE}(\nu_\mu)
\end{array}\right]       =
\left[\begin{array}{cc}
			1-P_E(\nu_2\to\nu_e) &        P_E(\nu_2\to\nu_e)  \\
            		P_E(\nu_2\to\nu_e) &            1-P_E(\nu_2\to\nu_e)
\end{array}\right]
\left[\begin{array}{c}
P_{S}(\nu_1)\\
P_{S}(\nu_2)
\end{array}\right]
\ ,  
\label{eq:PSE}
\end{equation}
where $P_E$ is the probability of the $\nu_2\to\nu_e$ transition
in the Earth. Equation~(\ref{eq:P_SE}) follows then from Eqs.~(\ref{eq:PS}) 
and (\ref{eq:PSE}).

	The incoherence of neutrino mass state components in Eqs.~(\ref{eq:PS})
and (\ref{eq:PSE}) is guaranteed by at least three facts:
(1) the neutrino production region in the Sun is an order of magnitude 
    larger than the Earth radius; 
(2) for typical values of neutrino  mass and mixing parameters, solar 
    neutrinos oscillate many times in their Sun-Earth path, with final
    wavepacket divergences larger than the oscillation wavelength; and 
(3) any detection process implies some energy smearing.
A hypothetical coherent mixture would give rather different numerical results 
for $P_{SE}$ \cite{Br89}.

\section{Proof of the property 
${\cal U}(0,\,-{\lowercase{x}})={\cal U}^T({\lowercase{x}},\,0)$}

	Let us consider the Schr{\"o}dinger equation 
\begin{equation}
	i\frac{d\psi(x)}{dx}={\cal H}(x)\psi(x)
\label{eq:Sch}
\end{equation}
and its formal solution
\begin{equation}
	\psi(x) = {\cal U}(x,\,0)\psi(0)
\ ,
\label{eq:psix}
\end{equation}
where $\cal U$ is the evolution operator (${\cal U}^\dagger{\cal U}=\openone$).

	If the Hamiltonian is real (${\cal H}={\cal H}^*$) and obeys the 
symmetry ${\cal H}(x)={\cal H}(-x)$, then $\psi^*(-x)$ is also a solution 
of Eq.~(\ref{eq:Sch}),
\begin{equation}
	\psi^*(-x) = {\cal U}(x,\,0)\psi^*(0)
\ ,
\label{eq:psi-x}
\end{equation}
that is
\begin{eqnarray}
		\psi(0) &=& [{\cal U}^*(x,\,0)]^{-1}\psi(-x)
\nonumber\\ 
        		&=&  {\cal U}^T(x,\,0)\psi(-x)
\ ,
\label{eq:psi}
\end{eqnarray}
which implies  that ${\cal U}(0,\,-x)={\cal U}^T(x,\,0)$.

\section{Changing integration measure,
$\int\!\lowercase{d}\tau_{\lowercase{d}}\int\!{\lowercase{d}}\tau_{\lowercase{h}} 
\to \int\! \lowercase{d}\eta$}

	In this appendix we show how to transform an integral of the
kind $\int\! d\tau_d\int\! d\tau_h$ into an integral of the kind $\int\!d\eta$. 
In particular,  Eq.~(\ref{eq:Pnight}) is explicitly derived
for a detector latitude between the Tropic and the Polar Circle
(see Table~\ref{tb:W}). Other cases are discussed  at the end of this 
appendix.

	The daily and hourly times are conventionally normalized to the 
interval $[0,\,2\pi]$:
\begin{eqnarray}
		\tau_d & = & \frac{\rm day}{365}\,2\pi
\ , \\
		\tau_h & = & \frac{\rm hour}{24}\,2\pi
\ ,
\end{eqnarray}
with $\tau_d=0$ at the winter solstice and $\tau_h=0$ at midnight.
The nadir angle $\eta$, the daily time $\tau_d$, and the hourly time $\tau_h$, 
are linked by the relations
\begin{eqnarray}
       \cos\eta &=&\cos\lambda\cos\tau_h\cos\delta_S-\sin\lambda\sin\delta_S\ ,
\label{eq:time}\\
    \sin\delta_S&=&-\sin i\cos\tau_d
\ ,
\label{eq:decl}
\end{eqnarray}
where $\lambda$ is the detector latitude (in radiants), $i$ is the Earth
inclination ($i=0.4091$~rad), and $\delta_S$ is the Sun declination.
The sunrise (sr) and sunset (ss) times (corresponding to $\eta=\pm\pi/2$)
are then given by $\tau_h^{\rm sr}=\arccos(\tan\lambda\tan\delta_S)$
and $\tau_h^{\rm ss}=-\tau_h^{\rm sr}$, respectively.

	The annual average during nights can be restricted, for symmetry,
to half year and to half night (midnight--sunrise interval),
\begin{equation}
\langle P_E \rangle_{\rm night}=
\frac{\displaystyle       \int^\pi_0 d\tau_d\int^{\tau_h^{\rm sr}(\tau_d)}_0
                               d\tau_h\,P_E(\eta(\tau_d,\,\tau_h))}
{\displaystyle            \int^\pi_0 d\tau_d\int^{\tau_h^{\rm sr}(\tau_d)}_0
                                             d\tau_h}
\ .
\label{eq:Phalf}
\end{equation}

	The integral at the denominator in Eq.~(\ref{eq:Phalf}) is trivial
and gives $\pi^2/2$. The integral at the numerator in Eq.~(\ref{eq:Phalf}) 
can be transformed as
\begin{eqnarray}\lefteqn{
\displaystyle 
		\int_0^{\pi}d\tau_d\int_{\lambda+\delta_S}^{\pi/2}
		d\eta\, \frac{d\tau_h}{d\eta}\,(\tau_d,\,\eta)\,P_E(\eta)}
\\&=&\displaystyle
		\int_{\lambda-i}^{\pi/2}d\eta\,P_E(\eta)\int_0^{\hat{\tau}_d
		(\eta)}d\tau_d\,\frac{d\tau_h}{d\eta}(\tau_d,\,\eta)
\label{eq:obscure}\\&=& 
		\frac{\pi^2}{2}\displaystyle\int_0^{\pi/2}d\eta\,P_E(\eta)
		\,W(\eta)
\ ,
\end{eqnarray}
where 
\begin{equation}
	\hat\tau_d(\eta)=\left\{\begin{array}{cl}
		0
					&,\;0\leq\eta<\lambda-i\ ,\\
\noalign{\smallskip}
		\arccos\left(\frac{\sin(\lambda-\eta)}{\sin i}\right) 
					&,\;\lambda-i\leq\eta<\lambda+i\ ,\\
\noalign{\smallskip}
		\pi 			&,\;\lambda+i\leq\eta\leq\pi/2
\ ,
\end{array}\right.
\label{eq:hat}
\end{equation}
and $W(\eta)$ is defined as
\begin{equation}
	W(\eta) = \frac{2}{\pi^2}\left\{\begin{array}{cl}
			0 	&,\;0\leq\eta<\lambda-i\ ,\\ 
\noalign{\smallskip}\displaystyle
	\int^{\hat\tau_d(\eta)}_0 d\tau_d\,\frac{d\tau_h}{d\eta}(\tau_d,\,\eta)
				&,\;\lambda-i\leq\eta\leq \pi/2
\ .
\end{array}\right.
\label{eq:WW}
\end{equation}

	The interchange of integration variables  in Eq.~(\ref{eq:obscure})
and the definition in Eq.~(\ref{eq:hat}) can be understood by drawing the 
integration domain in the $(\tau_d,\,\eta)$ plane (not shown) for a detector 
latitude between the Tropic and the Polar Circle $(i<\lambda\leq \pi/2-i)$.

	From Eqs.~(\ref{eq:time}) and (\ref{eq:decl}) one derives, after 
some algebra,
\begin{equation}
\displaystyle
	\int_0^{\hat\tau_d(\eta)}d\tau_d\frac{d\tau_h}{d\eta}(\tau_d,\,\eta)
	=
	\frac{\sin\eta}{\sin i}\int^1_{\max\{p,\,-1\}}
	\frac
						{d\xi}
				{\sqrt{(\xi+1)(\xi-1)(\xi-p)(\xi-q)}}
\ ,
\label{eq:preK}
\end{equation}
where
\begin{mathletters}
\begin{eqnarray}
 	p 	&=& 	\sin(\lambda-\eta)/\sin i\ ,\\
 	q 	&=& 	\sin(\lambda+\eta)/\sin i\ ,\\
	\xi	&=& 	\cos\tau_d
\ .
\end{eqnarray}
\end{mathletters}

	The  r.h.s.\ of Eq.~(\ref{eq:preK}) can be expressed in terms 
of the complete elliptic integral of the first kind, defined as 
\begin{equation}
	K(x)=\int_0^1 {\frac{ds}{\sqrt{(1-s^2)(1-x^2s^2)}}}\ .
\end{equation}
(see \cite{Gr94}, pp.~241--243). The results in the third column 
of Table~\ref{tb:W} are finally obtained with the positions
\begin{mathletters}
\begin{eqnarray}
		z	&=& 	\sin i\,\sqrt{(q-p)/2}\ ,\\
		y	&=& 	\sin i\,\sqrt{(1-p)(1+q)/4}
\ ,
\end{eqnarray}
\end{mathletters}
\noindent that can be easily shown to coincide with the definitions
in the bottom row of Table~\ref{tb:W}.

	The other cases reported in Table~\ref{tb:W} (nearly equatorial or 
polar detector latitudes) can be derived analogously, the only difference
being the shape of the integration domain in the $(\tau_d,\,\eta)$ 
rectangle. The above calculation can be specialized to annual averages 
during specific fractions of night, such as the period in which the Earth 
core is crossed [see Eq.~(\ref{eq:Pcore}) and related comments].

	As concerns the case in which the time average is taken over a 
fraction of year (e.g., a season), we only mention  that the weight function 
can still be expressed analytically, but the generic integration limits 
for $\tau_d$ require the calculation of the  {\em incomplete\/} elliptic 
integral of the first kind (see \cite{CERN} for its numerical evaluation).
The possible cases for the functional form of $W(\eta)$ acquire an additional 
dependence on the fraction of year considered for the average, and are 
not discussed in this paper.

\section{Effect of Earth-Sun distance variations}

	Throughout this work, the Earth-Sun distance $L$ has been taken 
constant $(L=L_0)$, in the hypothesis that the trivial $1/L^2$ geometrical 
variations of the solar neutrino signal will be factorized out in real-time 
experiments. However, such a continuous correction of the data
implies a real-time subtraction of the background and thus requires
a difficult, daily task of monitoring background, efficiencies, and 
calibrations (which are instead better defined over large periods of time).
Therefore, we consider also the effect of dropping the assumption
of a real-time, geometric correction of the signal, for the relevant case of
annual averages during (a fraction of) nighttime.

	Given the orbital equation 
\begin{equation}
	L(\tau_d)=L_0(1-\epsilon \cos(\tau_d-\tau_d^p)) 
		+ {\cal O}(\epsilon^2)
\ ,
\end{equation}
where $\tau_d^p=0.24$ corresponds to the perihelion and $\epsilon=0.0167$
is the Earth orbit eccentricity, the time-averaged probability reads
\begin{equation}
\langle P_E \rangle'_{\rm night}=
\frac{\displaystyle		\int^{2\pi}_0 d\tau_d\,L^{-2}(\tau_d)
				\int^{\tau_h^{\rm sr}(\tau_d)}_0
				d\tau_h\,P_E(\eta(\tau_d,\,\tau_h))}
{\displaystyle
				\int^{2\pi}_0 d\tau_d\, L_0^{-2}
				\int^{\tau_h^{\rm sr}(\tau_d)}_0
				d\tau_h}
\ .
\label{eq:P'half}
\end{equation}

We give without proof the final results for detector latitudes between the
Tropic and the Polar Circle (see also Table~\ref{tb:W}):
\begin{equation}
	\langle P_E \rangle_{\rm night}' = 
	\int^{\pi/2}_0 d\eta\,W'(\eta)\,P_E(\eta)
\ ,
\end{equation}
\begin{equation}
	W'(\eta) = W(\eta) \pm \epsilon\, Y(\eta)
\ ,
\end{equation}
where the upper (lower) sign refer to the northern (southern) hemisphere, 
$W(\eta)$ is given in Table~\ref{tb:W} (first and third column), and
the function $Y(\eta)$ is defined as
\begin{equation}
	Y(\eta) = 
	\frac{4}{\pi^2}\cos\tau_d^p\sin\eta\cdot\left\{
\begin{array}{cl}
	0 			& ,\;\ 0\leq\eta<\lambda-i\ , 
\\ \noalign{\bigskip}
	\frac{1}{z}\left[q\,K\left(\frac{y}{z}\right)-(q-1)\,
	\Pi\left(\frac{1-p}{q-p},\,\frac{y}{z}\right)\right]
				& ,\;\ \lambda-i\leq\eta<\lambda+i\ , 
\\ \noalign{\bigskip}
	\frac{1}{y}\left[q\,K\left(\frac{z}{y}\right)-(q-1)\,
	\Pi\left(\frac{2}{q+1},\,\frac{z}{y}\right)\right]
				& ,\;\ \lambda+i<\eta\leq\pi/2
\ . 
\label{eq:Pi}
\end{array}\right.
\end{equation}

	In Eq.~(\ref{eq:Pi}) the variables $p$, $q$, $y$, and $z$, are 
defined as in Appendix~C, and $\Pi$ is the complete elliptic integral of 
the third kind \cite{Gr94,Ab72},
\begin{equation}
	\Pi(r,\,x)=\displaystyle\int^1_0
	\frac{ds}{(1-r s^2)\sqrt{(1-s^2)(1-x^2 s^2)}}
\end{equation}
(see \cite{CERN} for its numerical evaluation).

The ``eccentricity correction'' $\pm \epsilon Y(\eta)$ is small. At 
latitudes of interest, the difference between annual averages with and 
without this term is less than 1\%:
\begin{equation}
   \left|\langle P_E\rangle'_{\rm night}-\langle P_E\rangle_{\rm night}\right|
   \leq \epsilon\int_0^{\pi/2}d\eta\,|Y(\eta)| =
   \left\{ \begin{array}{cl}
				0.82\% & {\rm\ (Kamioka)}\ ,
\\ \noalign{\medskip}
				0.95\% & {\rm\ (Sudbury)}\ ,
\\ \noalign{\medskip}
				0.90\% & {\rm\ (Gran\ Sasso)}
\ .
\end{array}
\right.
\end{equation}

	Finally, we mention that, for averages over fractions of year, 
the eccentricity correction involves the evaluation of {\em incomplete\/}
elliptic integrals of the third kind.


\begin{table}
\caption{	Coefficients of the electron density parametrization
		$N_j(r)=\alpha_j+\beta_jr^2+\gamma_jr^4$, $[N]={\rm mol/cm}^3$,
		for the $j$-th shell range $[r_{j-1},\,r_j]$. The radial 
		distance $r$ is normalized to the Earth radius. See 
		Fig.~\protect\ref{F:1} for a plot of $N(r)$.}
\smallskip
\begin{tabular}{clcddd}
$j$& Shell          & $[r_{j-1},\,r_j]$  &$\alpha_j$ &$\beta_j$ &$\gamma_j$\\
\tableline
1 & Inner core      &     $[0,\,0.192]$    &   6.099 & $-$4.119 &    0.000 \\
2 & Outer core      &   $[0.192,\,0.546]$  &   5.803 & $-$3.653 & $-$1.086 \\
3 & Lower mantle    &   $[0.546,\,0.895]$  &   3.156 & $-$1.459 &    0.280 \\
4 & Transition Zone &   $[0.895,\,0.937]$  &$-$5.376 &   19.210 &$-$12.520 \\
5 & Upper mantle    &     $[0.937,\,1]$    &  11.540 &$-$20.280 &   10.410 
\end{tabular}
\label{tb:N}
\end{table}

\vfil

\begin{table}
\caption{Weight function $W(\eta)$ for annual averages at the
	 latitude $\lambda$ ($i$ denotes the Earth inclination). 
	 $W(\eta)$ takes different functional forms in the indicated ranges 
	 of $\eta$ and $\lambda$.  $K$ is the complete elliptic function 
	 of the first kind, with arguments defined in the bottom row.}
\smallskip
\begin{tabular}{cccc} 
  Weight function   &  \multicolumn{3}{c}{
                Ranges of detector latitude $\lambda$ and nadir angle $\eta$}\\ 
\cline{2-4}
     $W(\eta)$ 
               &Equator to Tropic
                              &Tropic to Polar Circle
                                        &Polar Circle to Pole                \\
for annual averages
               &$(0\leq\lambda\leq i)$
                              &$(i<\lambda\leq\pi/2-i)$
                                        &$(\pi/2-i<\lambda\leq\pi/2)$     \\ 
\tableline  
\noalign{\medskip}
        0                         
                     & --- 
                              & $0\leq\eta<\lambda-i$ 
                                        & $0\leq\eta\leq\lambda-i$        \\ \\
$\displaystyle\frac{2\sin\eta}{\pi^2 z}\,K(y/z)$ 
                     &$i-\lambda<\eta<i+\lambda$ 
                              &  $\lambda-i\leq\eta<\lambda+i$ 
                                        &$\lambda-i\leq\eta<\pi-\lambda-i$\\ \\
$\displaystyle\frac{2\sin\eta}{\pi^2 y}\,K(z/y)$ 
                     & $0\leq\eta<i-\lambda$\ , 
                              & $\lambda+i<\eta\leq\pi/2$
                                        & $\pi-\lambda-i<\eta\leq\pi/2$      \\
\noalign{\vskip-4pt}
                     &or\ \ $i+\lambda<\eta\leq\pi/2$
                              & 
                                        &                                    \\
\noalign{\medskip}\tableline
\noalign{\smallskip}
Definitions:& \multicolumn{3}{l}{$z=\sqrt{\sin i\cos\lambda\sin\eta}$\ ,\ 
                                 $y=\sqrt{\sin\frac{i+\lambda+\eta}{2}
                                          \sin\frac{i-\lambda+\eta}{2}
                                          \cos\frac{i+\lambda-\eta}{2}
                                          \cos\frac{i-\lambda-\eta}{2}}$}
\end{tabular}
\label{tb:W}
\end{table}




\newcommand{\InsertFigure}[2]{\newpage\begin{center}\mbox{%
\epsfig{bbllx=1.4truecm,bblly=1.3truecm,bburx=19.5truecm,bbury=26.5truecm,%
height=21.truecm,figure=#1}}\end{center}\vspace*{-1.85truecm}%
\parbox[t]{\hsize}{\small\baselineskip=0.5truecm\hskip0.5truecm #2}}

\InsertFigure{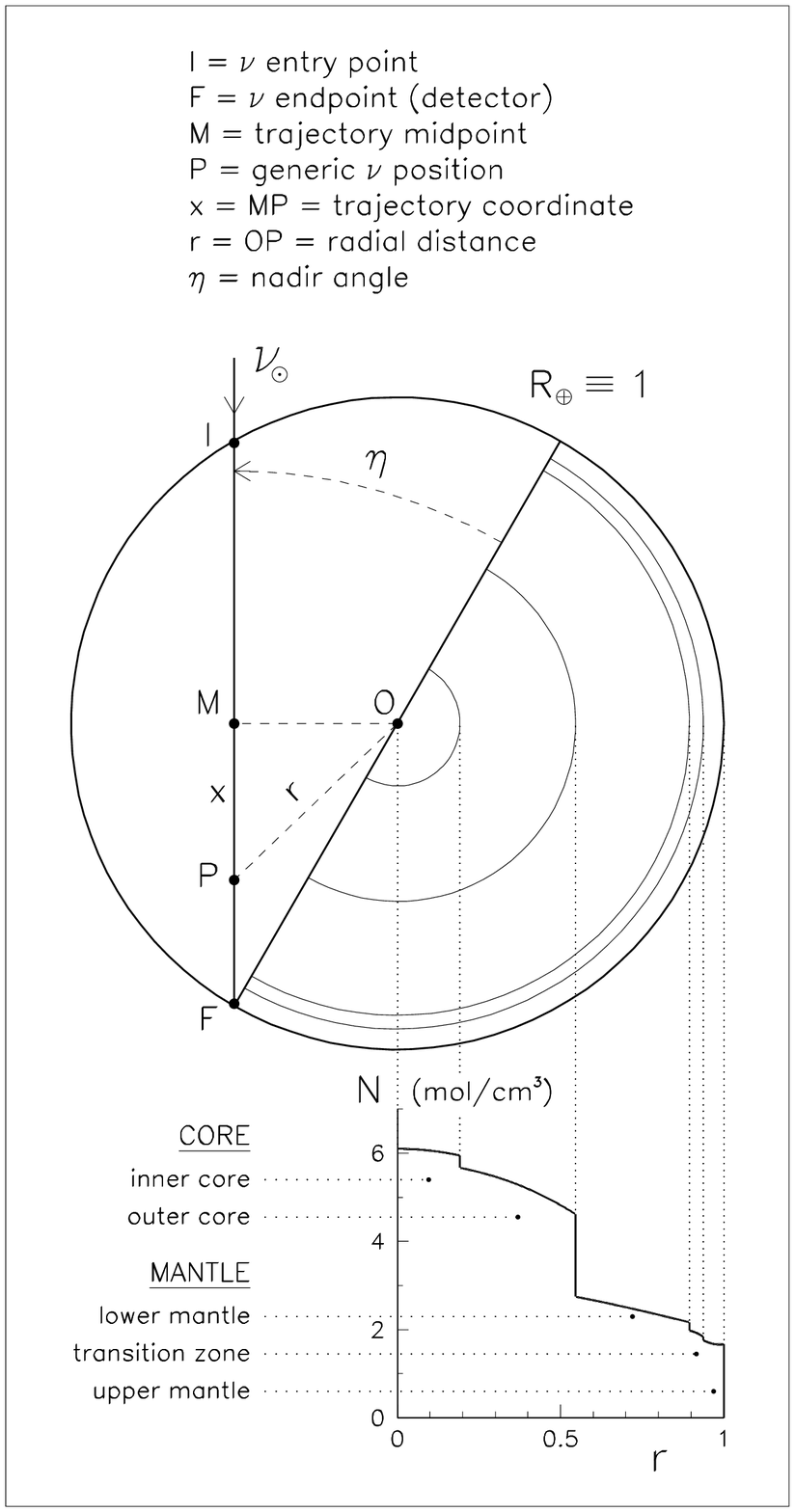}%
{FIG.~1.	Section of the Earth showing the relevant
		shells (in scale) and  the electron density profile 
		$N(r)$. The geometric definitions used in the text 
		are also displayed.}
\InsertFigure{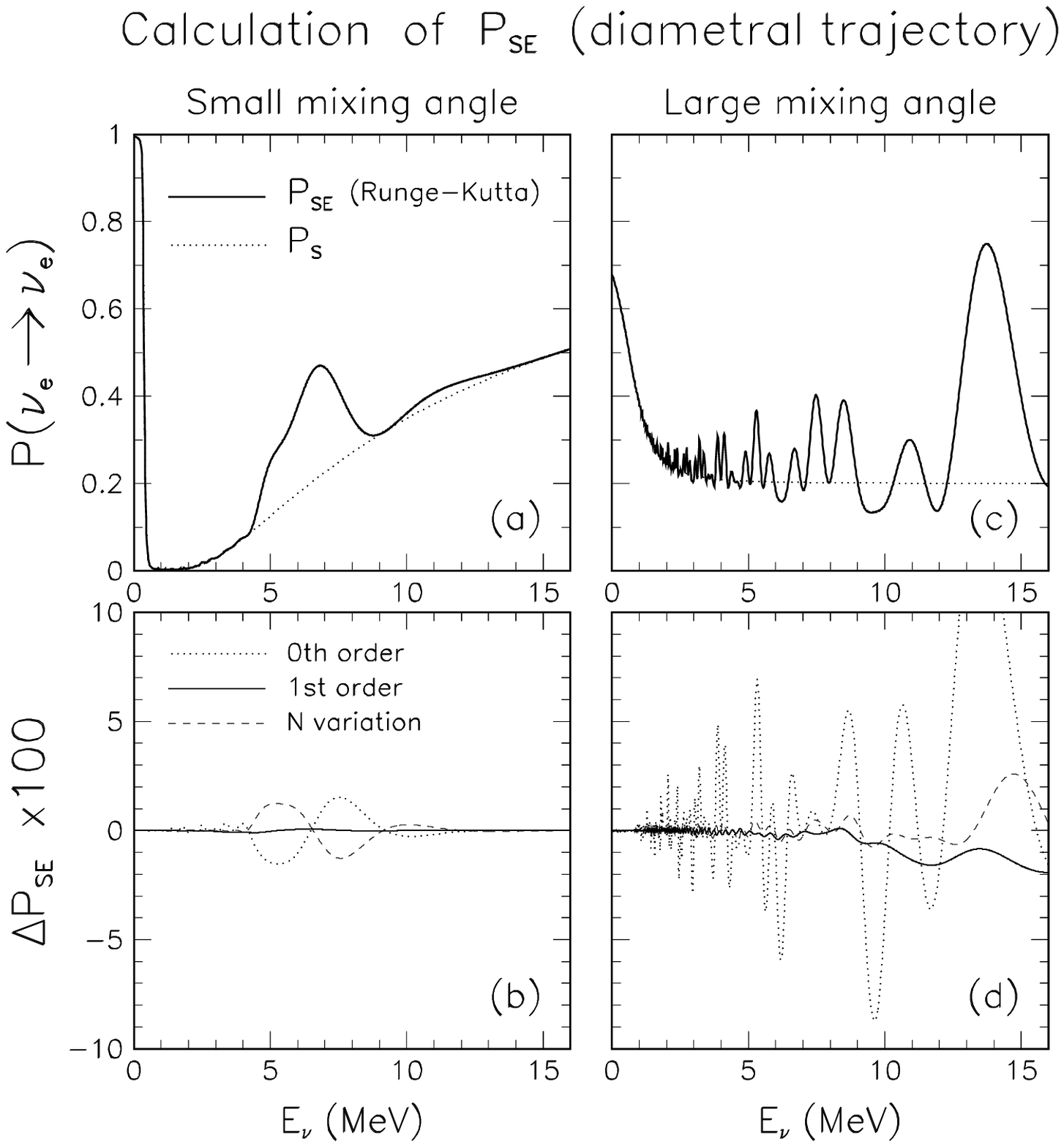}%
{FIG.~2.	Comparison of different calculations of $P_{SE}(E_\nu)$ for
		$^8$B neutrinos crossing the Earth diameter.
		(a) Calculation of $P_{SE}$ with Runge-Kutta integration for
		the small mixing angle case (solid line). 
		Also shown is the function $P_S$ (dotted line).
		(b) Variations of $P_{SE}$ induced by representative
		density shifts (dashed line), and by the first-order
		and zeroth order approximations discussed in the text
		(solid and dotted line, respectively). Panels (c) and (d)
		are analogous to (a) and (b), but refer to the large mixing
		angle case.}
\InsertFigure{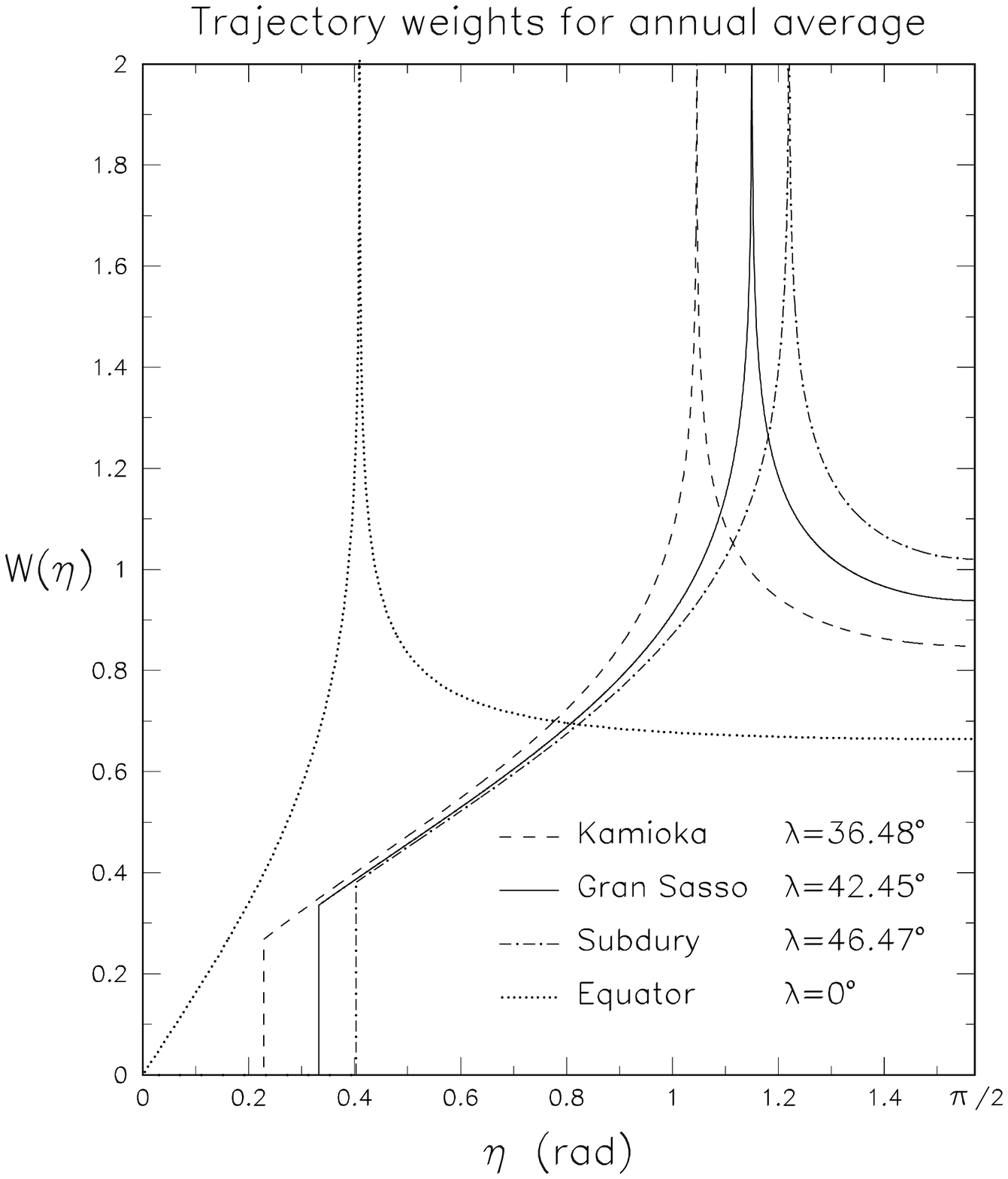}%
{FIG.~3.	Annual solar exposure (weight) of the trajectory
		at nadir angle $\eta$ for representative values
		of the  latitude $\lambda$. See the text for details.}
\InsertFigure{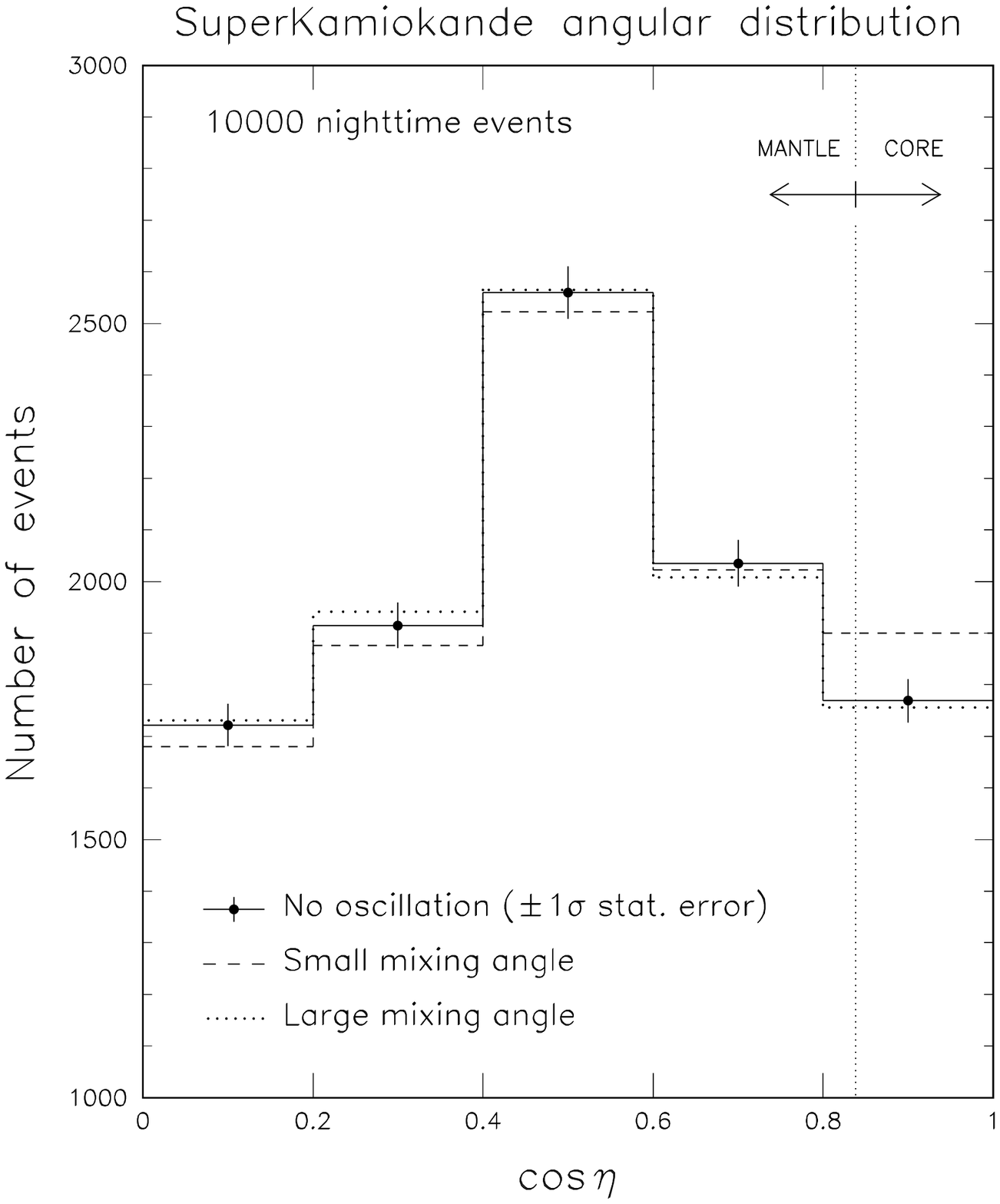}%
{FIG.~4.	Nadir angle distribution of nighttime events at
		SuperKamiokande. Error bars are statistical
		only.}
\InsertFigure{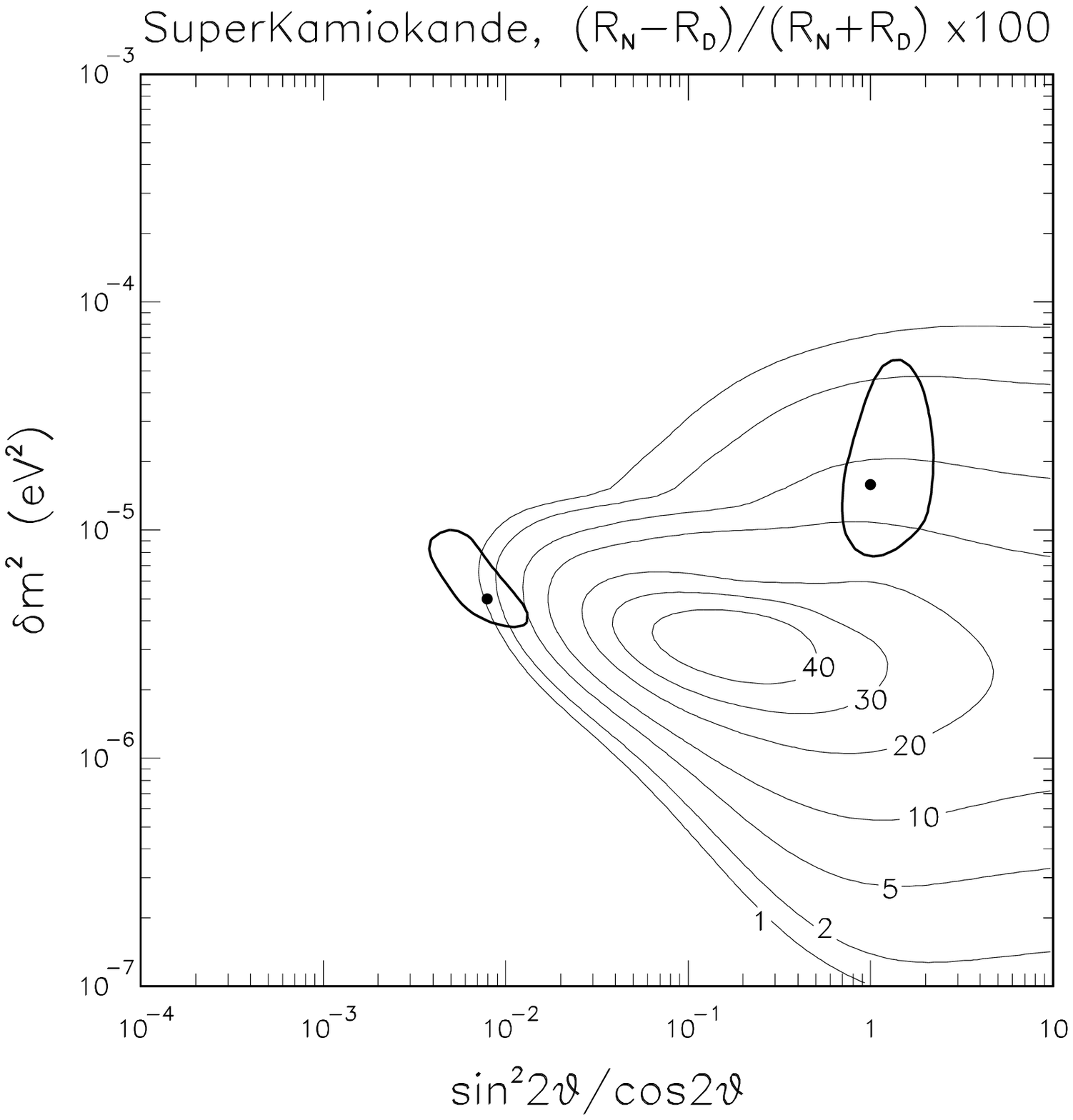}%
{FIG.~5.	Night-day asymmetry of neutrino rates at SuperKamiokande.
		The best-fit points and the 90\% C.L.\ regions
		for the small and large mixing angle solutions are
		superposed.}
\InsertFigure{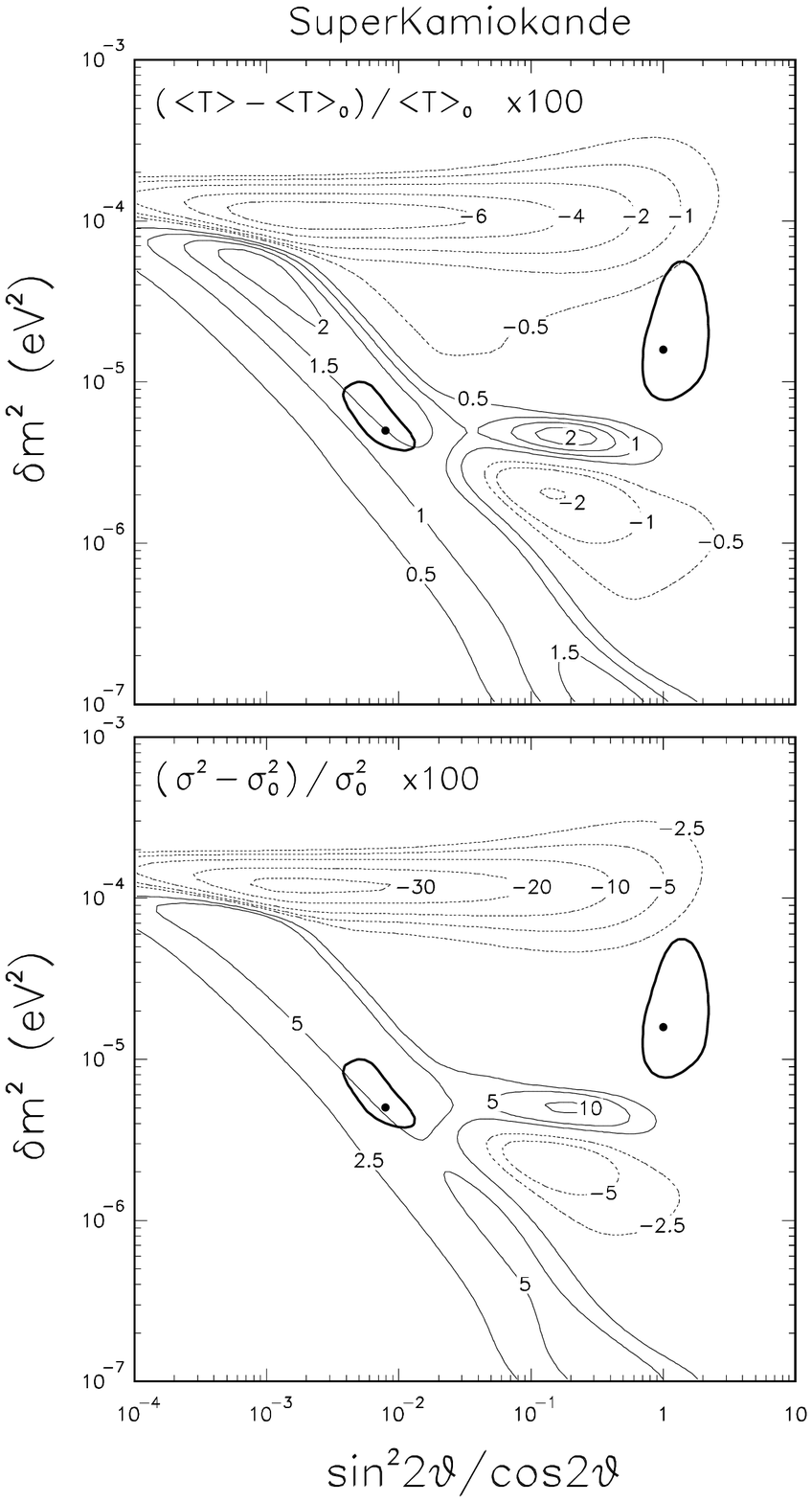}%
{FIG.~6.	Fractional deviations of the first two moments of
		the SuperKamiokande
		electron energy distribution ($\langle T\rangle$
		and $\sigma^2$) from their no-oscillation values
		($\langle T\rangle_0$
		and $\sigma^2_0$).}
\InsertFigure{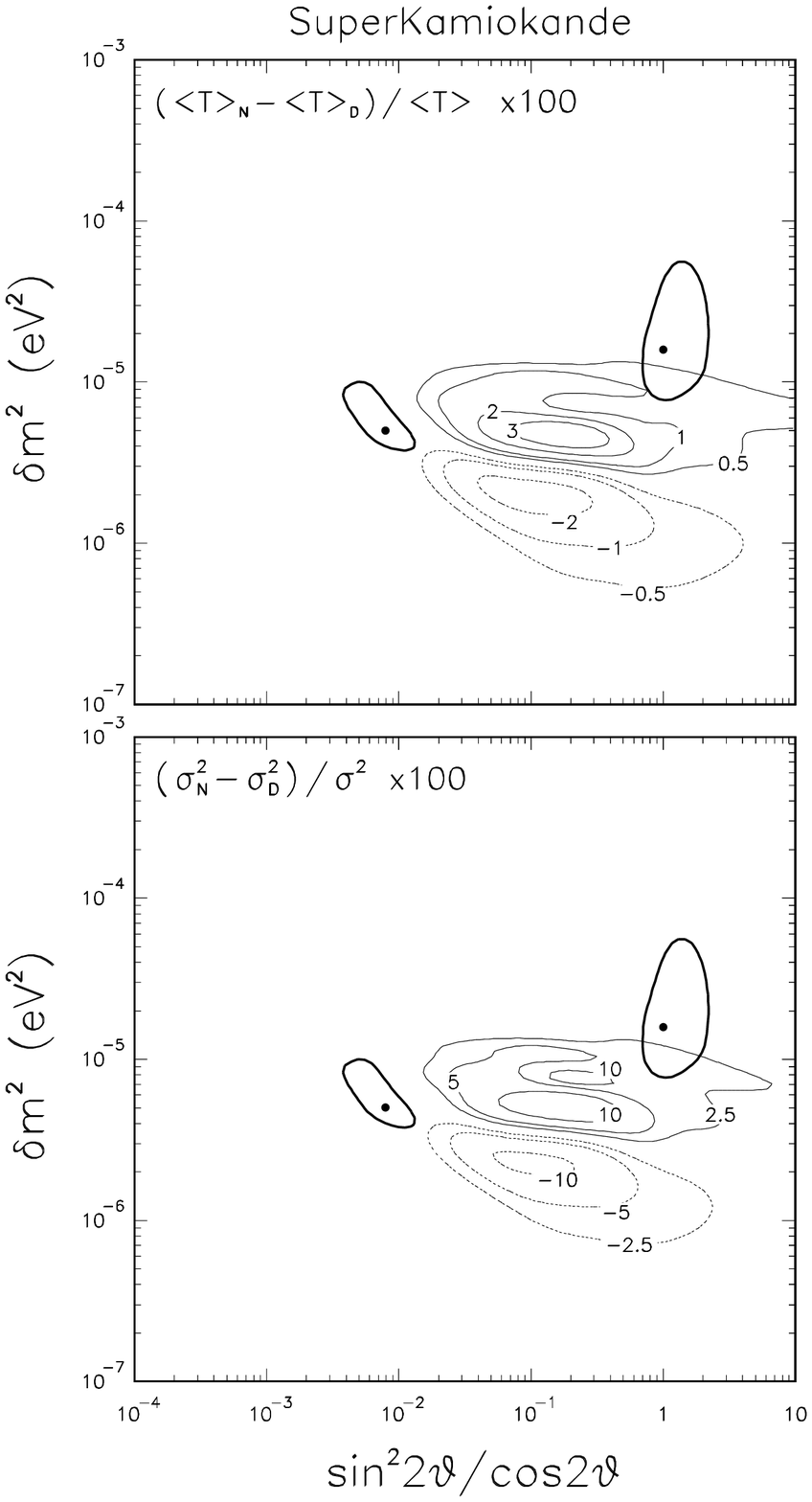}%
{\hfil FIG.~7.	Night-day fractional
		variations of the spectral moments at 
		SuperKamiokande.\hfil}
\InsertFigure{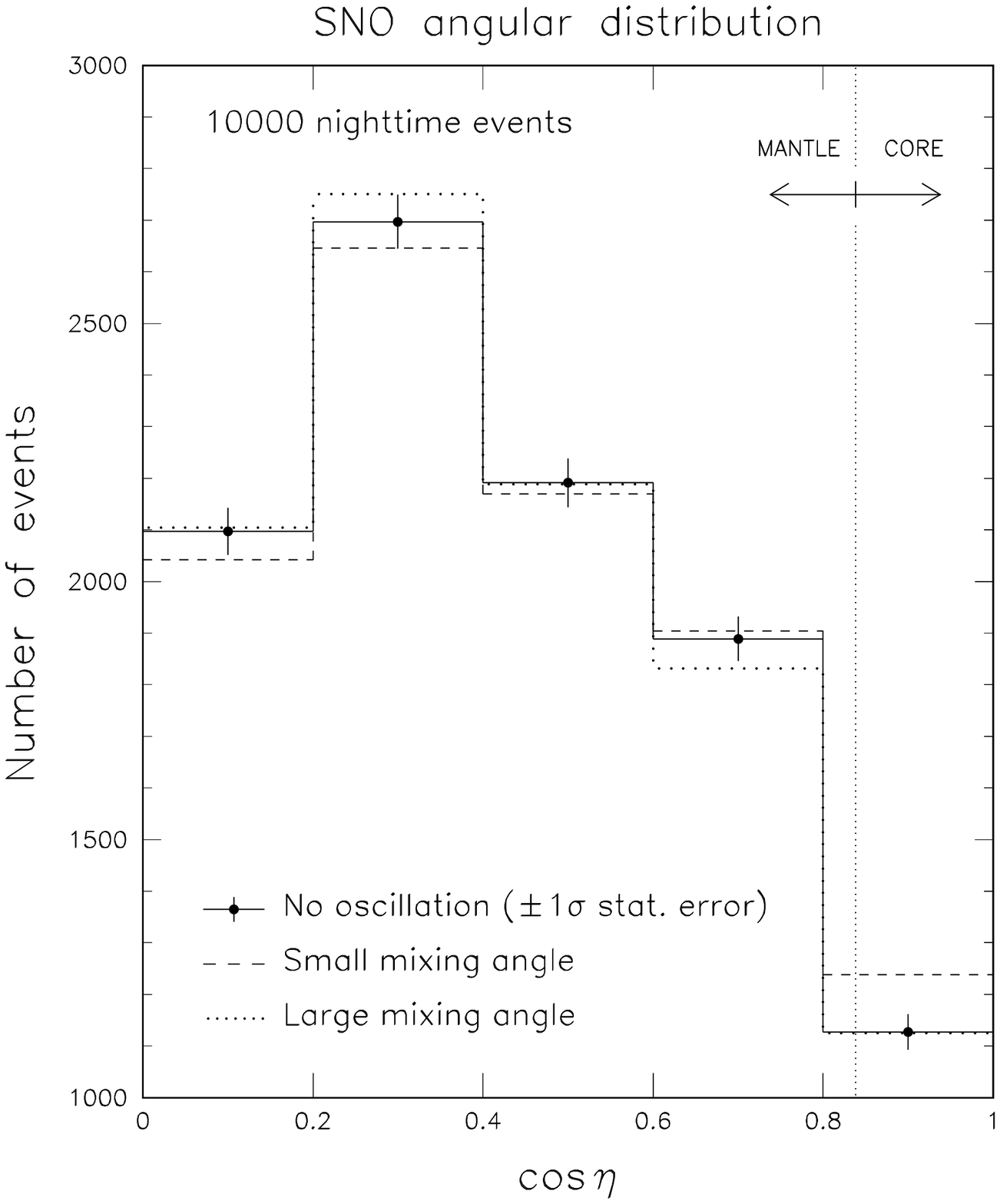}%
{FIG.~8.	Nadir angle distribution of nighttime events at
		SNO. Error bars are statistical
		only.}
\InsertFigure{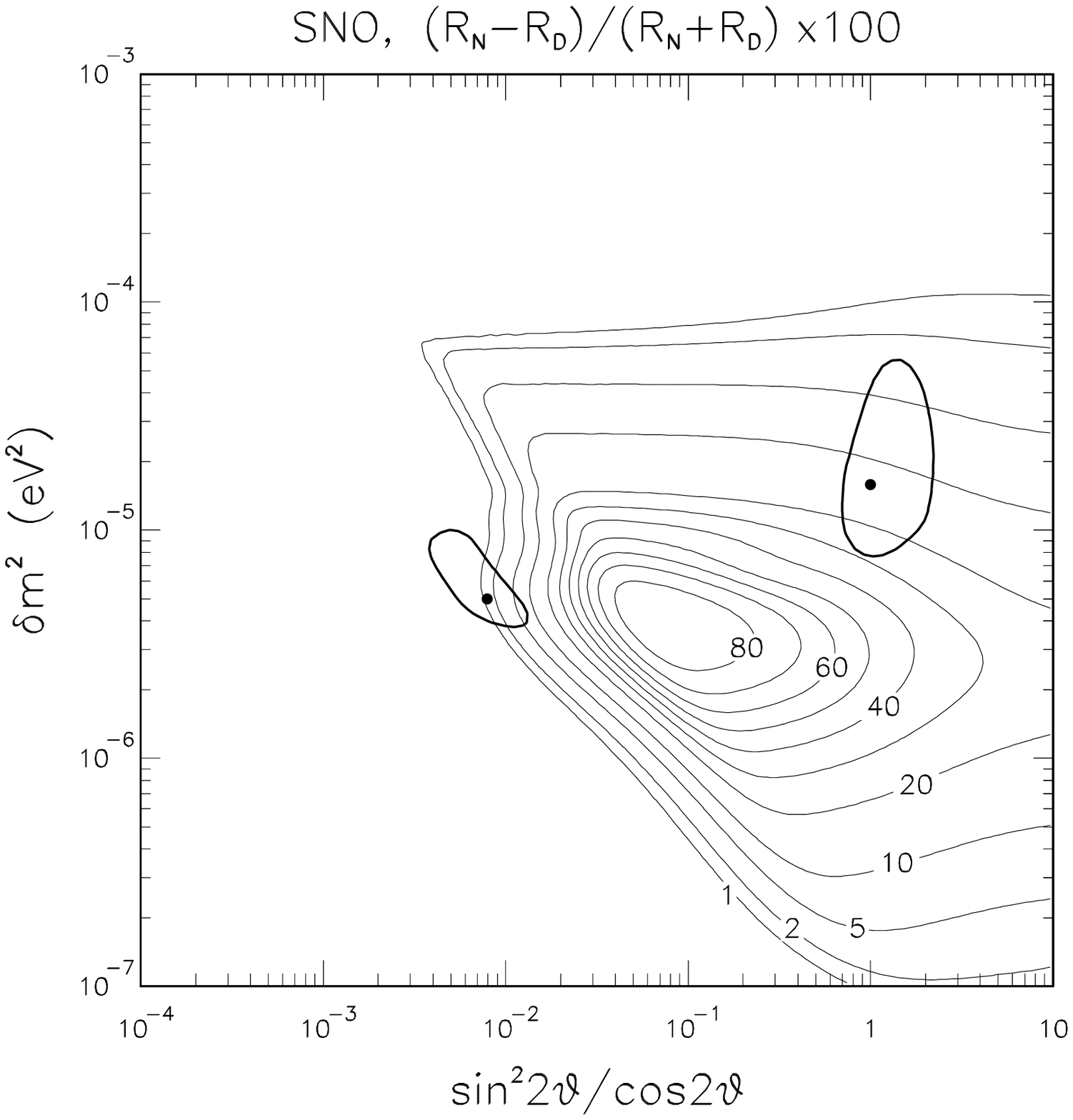}%
{\hfil FIG.~9.	Night-day asymmetry of neutrino rates at SNO.\hfil }
\InsertFigure{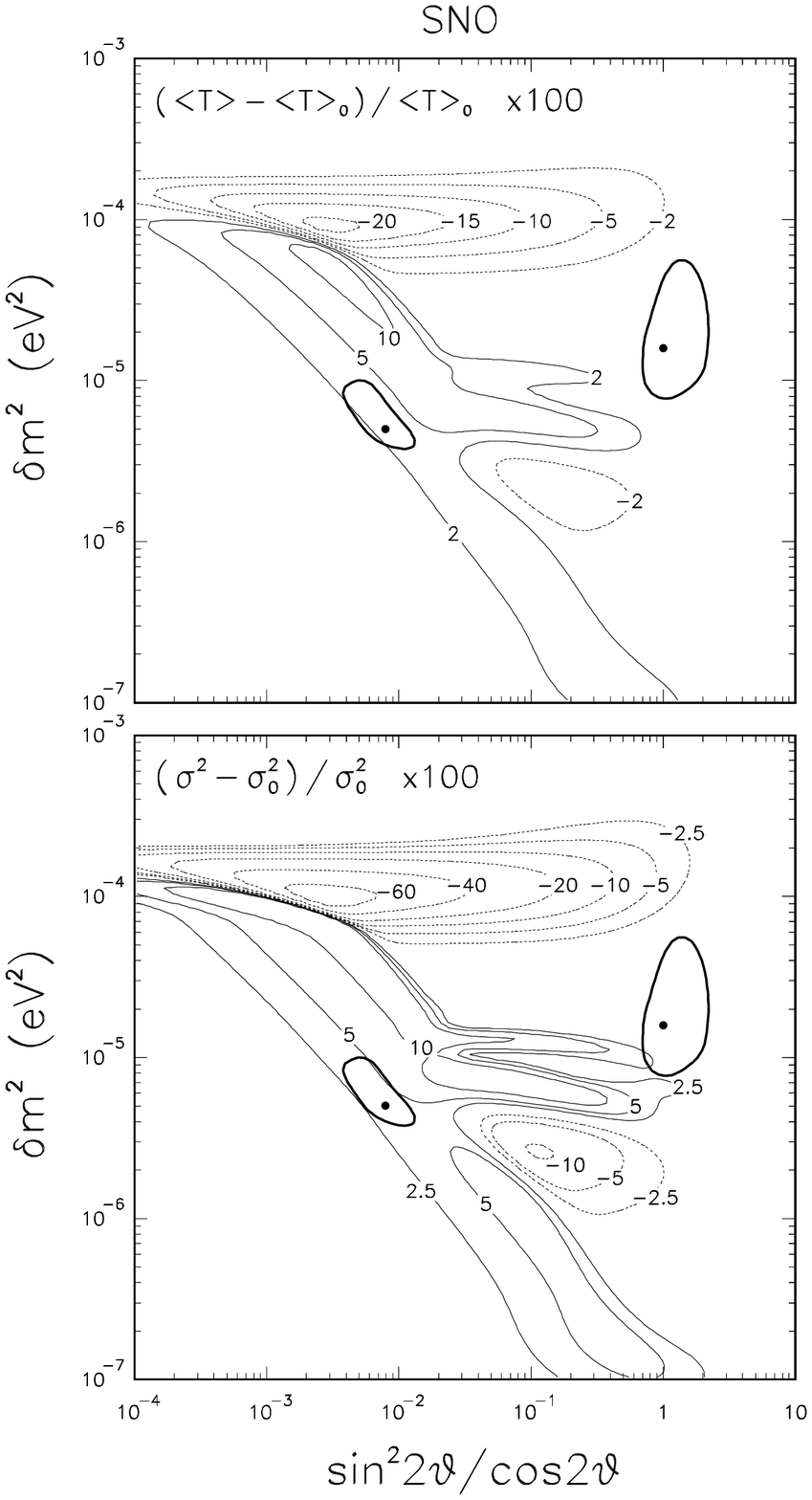}%
{FIG.~10.	Fractional deviations of the first two moments of
		the SNO electron energy distribution ($\langle T\rangle$
		and $\sigma^2$) from their no-oscillation values.}
\InsertFigure{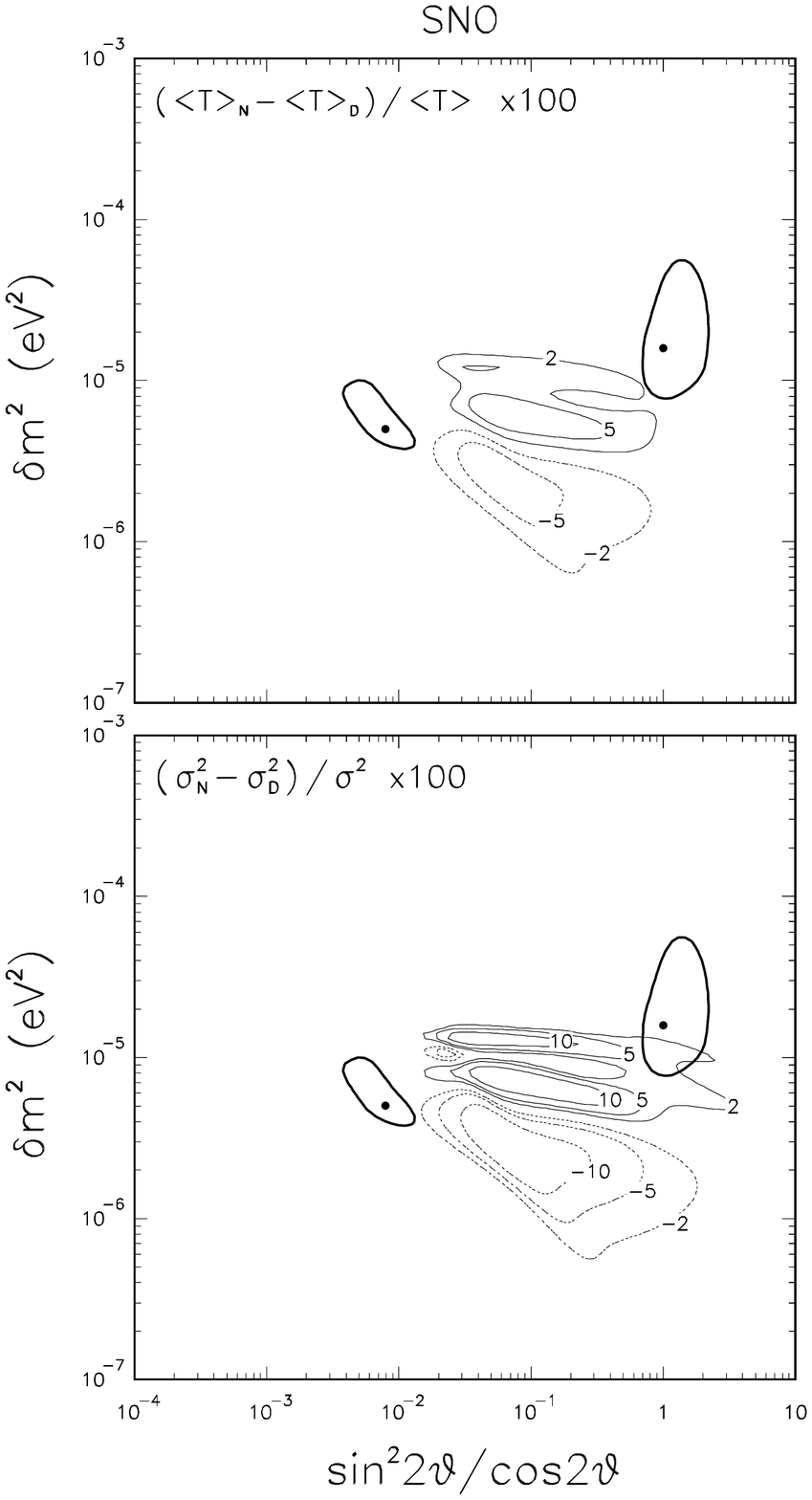}%
{\hfil FIG.~11.	Night-day fractional
		variation of the spectral moments at SNO.\hfil }
\InsertFigure{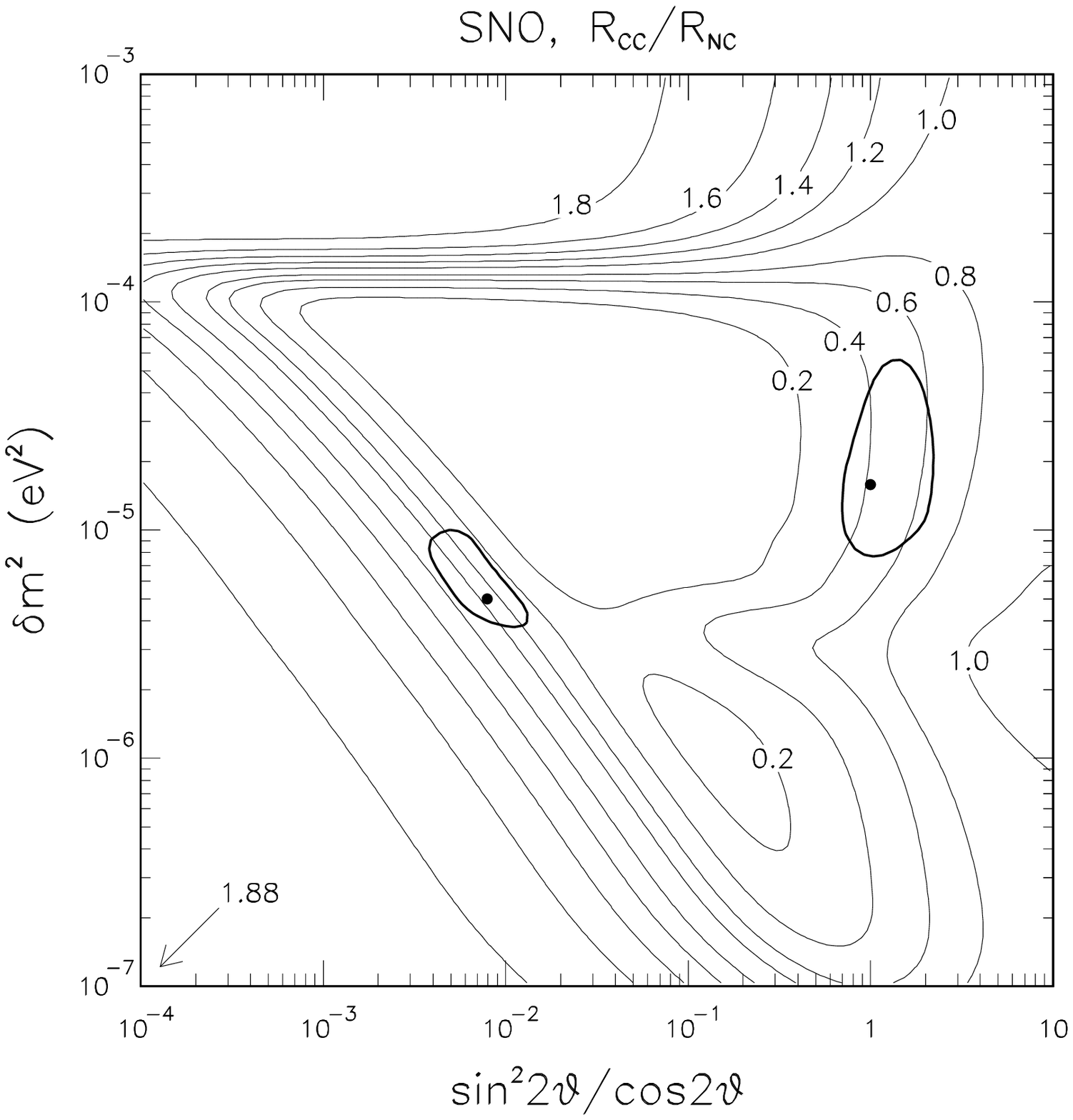}%
{\hfil FIG.~12.	Ratio of charged current to neutral current neutrino
		interactions at SNO.\hfil}


\begin{thebibliography}{99}
\bigskip\bigskip

\bibitem{MSWm}	L.\ Wolfenstein, 
		Phys.\ Rev.\ D {\bf 17}, 2369 (1978);
		S.\ P.\ Mikheyev and A.\ Yu.\ Smirnov, 
		Yad.\ Fiz.\ {\bf 42}, 1441 (1985) 
		[Sov.\ J.\ Nucl.\ Phys.\ {\bf 42}, 913 (1985)];
		Nuovo Cim. C {\bf 9} (1986), 17.

\bibitem{NuAs}	J.\ N.\ Bahcall, {\em Neutrino Astrophysics}
		(Cambridge University Press, Cambridge, England, 1989).

\bibitem{olds}	J.\ Bouchez, M.\ Cribier, W.\ Hampel, J.\ Rich, and D.\ Vignaud,
		Z.\ Phys.\ C {\bf 32}, 499 (1986);
		M.\ Cribier, W.\ Hampel, J.\ Rich, and D.\ Vignaud,
		Phys.\ Lett. B {\bf 182}, 89 (1986);
		A.\ J.\ Baltz and J.\ Weneser,
		Phys.\ Rev.\ D {\bf 35}, 528 (1987); {\bf 37}, 3364 (1988). 
		A.\ Dar, A.\ Mann, Y.\ Melina, and D.\ Zajfman,
		Phys.\ Rev.\ D {\bf 35}, 3607 (1987);
		A.\ Dar and A.\ Mann, 
		Nature (London) {\bf 325}, 790 (1987);
		S.\ P.\ Mikheyev and A.\ Yu.\ Smirnov,
		in {\em Moriond '87}, Proceedings of the 7th Moriond Workshop
		on New and Exotic Phenomena, Les Arcs, 1987, edited by
		O.\ Fackler and J.\ Tr{\^{a}}n Thanh V{\^{a}}n 
		(Fronti{\`e}res, Paris, 1987), p.\ 405;
		M.\ L.\ Cherry and K.\ Lande,
		Phys.\ Rev.\ D {\bf 36}, 3571 (1987);
		S.\ Hiroi, H.\ Sakuma, T.\ Yanagida, and M.\ Yoshimura,
		Phys.\ Lett.\ B {\bf 198}, 403 (1987); 
		Prog.\ Theor.\ Phys.\ {\bf 78}, 1428 (1987);
		M.\ Spiro and D.\ Vignaud,
		Phys.\ Lett.\ B {\bf 242}, 279 (1990).

\bibitem{Ku89}	T.\ K.\ Kuo and J.\ Pantaleone,
		Rev.\ Mod.\ Phys.\ {\bf 61}, 937 (1989).

\bibitem{Mi89}	S.\ P.\ Mikheyev and A.\ Yu.\ Smirnov, 
		Prog.\ Part.\ Nucl.\ Phys.\ {\bf 23}, 41 (1989).

\bibitem{Hi91}	Kamiokande Collaboration, K.\ S.\ Hirata {\em et al.},
		Phys.\ Rev.\ Lett.\ {\bf 66}, 9 (1991).

\bibitem{Fu96}	Kamiokande Collaboration, Y.\ Fukuda {\em et al.},
		Phys.\ Rev.\ Lett.\ {\bf 77}, 1683 (1996).

\bibitem{Ha93}	N.\ Hata and P.\ Langacker,
		Phys.\ Rev.\ D {\bf 48}, 2937 (1993); {\bf 50}, 632 (1994).

\bibitem{Fi94}	G.\ Fiorentini, M.\ Lissia, G.\ Mezzorani, M.\ Moretti, 
		and D.\ Vignaud,
		Phys.\ Rev.\ D {\bf 49}, 6298 (1994).

\bibitem{Fo96}	G.\ L.\ Fogli, E.\ Lisi, and D.\ Montanino,
		Phys.\ Rev.\ D {\bf 49}, 3626 (1994); 
		Astropart.\ Phys.\ {\bf 4}, 177 (1995);
		Phys.\ Rev.\ D {\bf 96}, 2048 (1996).

\bibitem{Ba94}	A.\ J.\ Baltz and J.\ Weneser,
		Phys.\ Rev. D {\bf 50}, 5971 (1994);
		{\bf 51}, 3960 (1995).

\bibitem{Kr96}  P.\ I.\ Krastev, in {\em DPF '96}, Proceedings of the
		1996 Annual Divisional Meeting of the Division of Particles 
		and Fields of the American Physical Society, Minneapolis,
		MN, 1996, to appear.

\bibitem{To95}	Y.\ Totsuka, in {\em TAUP '95}, Proceedings of the
		4th International Workshop on Theoretical and Phenomenological
		Aspects of Underground Physics, Toledo, Spain, edited by
		A.\ Morales, J.\ Morales, and J.\ A.\ Villar 
		[Nucl.\ Phys.\ B (Proc.\ Suppl.) {\bf 48}, 547
		(1996)]; A.\ Suzuki, in {\em Physics and Astrophysics
		of Neutrinos}, edited by M.\ Fukugita and A.\ Suzuki
		(Springer-Verlag, Tokyo, 1994), p.\ 414.

\bibitem{Do96}	A.\ B.\ McDonald, Proceedings of the 9th
		Lake Louise Winter Institute, edited by
		A.\ Astbury {\em et al.}, (World Scientific,
		Singapore, 1994), p.~1; {\em TAUP '95} \cite{To95},
		p.\ 357.

\bibitem{Ga95}	E.\ Gates, L.\ M.\ Krauss, and M.\ White,
		Phys.\ Rev.\ D {\bf 51}, 2631 (1995).

\bibitem{PREM}	A.\ M.\ Dziewonski and D.\ L.\ Anderson,
		Phys.\ Earth Planet.\ Inter.\ {\bf 25}, 297 (1981).

\bibitem{Re95} 	{\em Properties of the Solid Earth},
		in Reviews of Geophysics {\bf 33} (1995); also available
		at the URL: http://earth.agu.org/revgeophys/~.

\bibitem{ASPH}	A.\ M.\ Dziewonski and J.\ H.\ Woodhouse,
		Science {\bf 236}, 37 (1987).

\bibitem{MANT}	D.\ L.\ Anderson, Science {\bf 243}, 367 (1989).

\bibitem{CORE}	D.\ J.\ Stevenson, Science {\bf 214}, 214 (1981).

\bibitem{Bi52}	F.\ Birch, 
		J.\ Geophys.\ Res.\ {\bf57}, 227 (1952); on p.~234:
		``Unwary readers should take warning that ordinary
		language undergoes modification to a high pressure
		form when applied to the interior of the earth, e.g.,		
		\par\begin{center}\begin{tabular}[c]{lcl}
		\tableline
		High pressure form      &\ \ \ \ \ \ \ \ \ \ \ \ & 
		                         Ordinary meaning 	 \\
		\tableline
		Certain			&& Dubious		 \\
		Undoubtedly		&& Perhaps		 \\
		Positive proof		&& Vague suggestion	 \\
		Unanswerable argument	&& Trivial objection	 \\
		Pure iron		&& Uncertain mixture of  \\	
					&& \ \ all the elements''\\
		\tableline\end{tabular}\end{center}

\bibitem{Ba95}	J.\ N.\ Bahcall and M.\ H.\ Pinsonneault,
		Rev.\ Mod.\ Phys.\ {\bf 67}, 781 (1995).

\bibitem{Gr94}	I.\ S.\ Gradshteyn and I.\ M.\ Ryzhik,
		{\em Tables of Integrals, Series, and Products\/}
		(Academic Press, San Diego CA, 1994).

\bibitem{Ab72}	M.\ Abramowitz and I.\ A.\ Stegun,
		{\em Handbook of Mathematical Functions with Formulas,
		Graphs, and Mathematical Tables\/} 
		(John Wiley and Sons, New York, 1972).

\bibitem{CERN}	CERN Program Library CERNLIB, subroutine packages
		C346 and C347 for the calculation of incomplete
		and complete elliptic integrals of the first, second,
		and third kind.  
		A printable description of these routines can be found at the
		URL http://wwwcn.cern.ch/asdoc/cernlib.html~.

\bibitem{BORE}	Borexino Collaboration, G.\ Bellini {\em et al.},
		in {\em TAUP~'95} \cite{To95}, p.~363.

\bibitem{ICAR}	ICARUS Collaboration, C.\ Rubbia {\em et al.},
		in {\em TAUP~'95} \cite{To95}, p.~172.

\bibitem{PGNO}	GNO Collaboration, ``Proposal for a permanent Gallium
		Neutrino Observatory (GNO) at Laboratori Nazionali del Gran
		Sasso'' (unpublished). The proposal is available at
		the URL http://kosmopc.mpi-hd.mpg.de/gallex/gallex.htm~.

\bibitem{HELZ}	HELLAZ Collaboration, T.\ Ypsilantis {\em et al.}, 
		``HELLAZ: a high rate solar neutrino detector with
		neutrino energy determination,'' College de France
		Report LPC/94-28 (unpublished).
		
\bibitem{EQUA}	J.\ M.\ LoSecco, 
		Phys.\ Rev.\ D {\bf 47}, 2032 (1993);
		J.\ M.\ Gelb, W.\ Kwong, and S.\ P.\ Rosen, University
		of Texas at Arlington Report No.\ hep-ph/9612332 (unpublished).

\bibitem{QUAD}	P.\ J.\ Davis and P.\ Rabinowitz,
		{\em Methods of Numerical Integration}
		(Academic Press, San Diego, CA, 1984).

\bibitem{Al96}	J.\ N.\ Bahcall, E.\ Lisi, D.\ E.\ Alburger,
		L.\ DeBraeckeleer, S.\ J.\ Freedman, and J.\ Napolitano,
		Phys.\ Rev.\ C {\bf 54}, 411 (1996).


\bibitem{Ba96}	J.\ N.\ Bahcall, P.\ I.\ Krastev, and E.\ Lisi,
		Phys. Rev.\ C {\bf 55}, 494 (1997); see also
		J.\ N.\ Bahcall and E.\ Lisi,
		Phys.\ Rev.\ D {\bf 54}, 5417 (1996).

\bibitem{TABL}	Tables of neutrino cross sections 
		and moments of electron spectra, available at the URL
		http://www.sns.ias.edu/$^\sim$jnb (see Neutrino Export
		Software and Data).

\bibitem{Br89}	D.\ Bruss and L.\ M.\ Sehgal,
		Phys.\ Lett.\ B {\bf 216}, 426 (1989).



\end{thebibliography}
\end{document}